\pdfoutput=1

 \documentclass[twocolumn]{aastex61}

\newcommand\aastex{AAS\TeX}

\shorttitle{\aastex\ The build-up history of Fornax A}
\shortauthors{Iodice et al.}

\begin{document}

\title{The Fornax Deep Survey with VST. II.  Fornax A: a two-phase assembly caught on act}

\correspondingauthor{Enrichetta Iodice}
\email{iodice@na.astro.it}

\author[0000-0003-4291-0005]{E. Iodice}
\affil{INAF-Astronomical Observatory of Capodimonte, via Moiariello 16, Naples, I-80131, Italy}

\author{M. Spavone}
\affiliation{Astronomical Observatory of Capodimonte, via Moiariello 16, Naples, I-80131, Italy}

\author{M. Capaccioli}
\affiliation{University of Naples ``Federico II'', C.U. Monte SantÕAngelo, Via Cinthia, 80126, Naples, Italy}

\author{R. F. Peletier}
\affiliation{Kapteyn Astronomical Institute, University of Groningen, PO Box 72, 9700 AV Groningen, The Netherlands}

\author{T. Richtler}
\affiliation{Universidad de Concepci{\'o}n, Concepci{\'o}n, Chile}

\author{M. Hilker}
\affiliation{European Southern Observatory, Karl-Schwarzschild-Strasse 2, D-85748 Garching bei Munchen, Germany}

\author{S. Mieske}
\affiliation{European Southern Observatory, Alonso de Cordova 3107, Vitacura, Santiago, Chile}

\author{L. Limatola, A. Grado, N.R. Napolitano}
\affiliation{INAF - Astronomical Observatory of Capodimonte, via Moiariello 16, Naples, I-80131, Italy}



\author{M. Cantiello}
\affiliation{INAF-Astronomical Observatory of Teramo, Via Maggini, 64100, Teramo, Italy}

\author{R. D'Abrusco}
\affiliation{Smithsonian Astrophysical Observatory/Chandra X-ray centre, 02138 Cambridge (MA), US}

\author{M. Paolillo}
\affiliation{Univ. of Naples ``Federico II'', C.U. Monte SantÕAngelo, Via Cinthia, 80126, Naples, Italy}

\author{A. Venhola}
\affiliation{Division of Astronomy, Department of Physics, University of Oulu, Oulu, Finland}

\author{T. Lisker}
\affiliation{Zentrum fuer Astronomie der Universitaet Heidelberg, Germany}

\author{G. Van de Ven}
\affiliation{Max Planck Institute for Astronomy, Heidelberg, Germany}

\author{J. Falcon-Barroso}
\affiliation{Instituto de Astrof\'isica de Canarias, C/ Via L\'actea s/n, 38200 La Laguna, Canary Islands, Spain}

\author{P. Schipani}
\affiliation{Astron. Observatory of Capodimonte, via Moiariello 16, Naples, I-80131, Italy}



\begin{abstract}

{  As part of the Fornax Deep Survey with the ESO VLT Survey Telescope,
we present new $g$ and $r$ bands mosaics of the SW group of the Fornax cluster. It covers an area of  $3 \times 2$~square degrees around the central galaxy NGC~1316.
The deep photometry, the high spatial resolution of OmegaCam and the large covered area allow us to study the galaxy structure, to trace stellar halo formation and look at the galaxy environment. 
We map the surface brightness profile out to  33~arcmin ($\sim 200$~kpc $\sim15R_e$) from the galaxy centre, down to $\mu_g \sim 31$~mag arcsec$^{-2}$  and $\mu_r \sim 29$~mag arcsec$^{-2}$. This allow us to estimate the scales of the main components dominating the light distribution, which are the central spheroid, inside 5.5~arcmin ($\sim33$~kpc), and the outer stellar envelope. 
Data analysis suggests that we are catching in act the second phase of the mass assembly in this galaxy, since the accretion of smaller satellites is going on in both components. The outer envelope of NGC~1316 still hosts the remnants of the accreted satellite galaxies that are forming the stellar halo.
We discuss the possible formation scenarios for NGC~1316, by comparing the  observed properties (morphology, colors, gas content, kinematics and dynamics)  with predictions from cosmological simulations of galaxy formation. We find that {\it i)} the central spheroid could result from at least one merging event, it could be a pre-existing early-type disk galaxy with a lower mass companion, and {\it ii)} the stellar envelope comes from the gradual accretion of small satellites.}

\end{abstract}





\section{Introduction}\label{intro}

\begin{figure*}[t]
\centering
\includegraphics[scale=0.7]{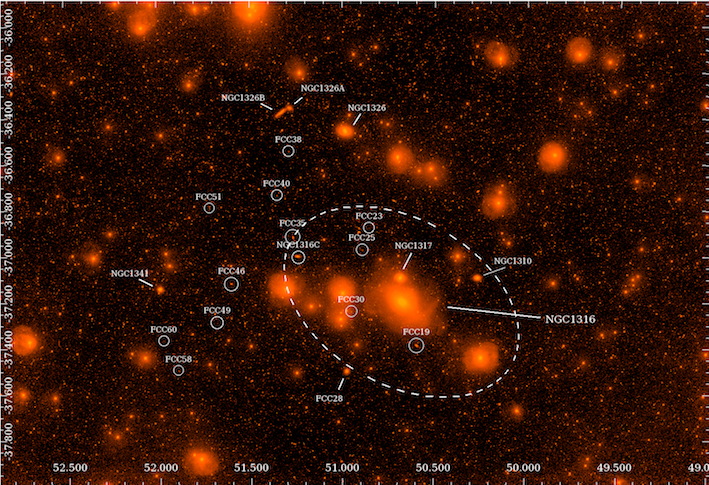}
\caption{\label{mosaic} Sky-subtracted VST mosaic in the $r$  band of the central $3.6 \times 2.1$~degrees$^2$ around NGC~1316, which the brightest object in the centre. All the galaxy members of this subgroup of the Fornax cluster \citep{Ferguson1989} are also indicated, including the dwarfs candidates (inside the white circles). Right Ascension and Declination (in degrees) are reported on the horizontal and vertical axes, respectively. }
\end{figure*}

{  
The hierarchical structure formation at all scales is one of the strongest prediction of the $\Lambda$CDM model. In this framework, cluster of galaxies are expected to grow over time by accreting smaller groups. The galaxies at the centre of the clusters continue to undergo active mass assembly and, in this process, gravitational interactions and merging among systems of comparable mass and/or smaller objects play a fundamental role in defining the observed structures. 
Since time scales in the outer parts are long, the imprints of mass assembly that we are currently observing are seen there:
 these are the regions of the stellar halo. This is an extended, diffuse and very faint ($\mu \sim 26-27$~mag/arcsec$^2$ in the $g$ band) component made of stars stripped from satellite galaxies, in the form of streams and tidal tails, with multiple stellar components and complex kinematics, which is still growing at the present epoch.
On the scale of the cluster, during the infall of groups of galaxies to form the cluster, the material stripped from the galaxy outskirts builds up the intracluster light, ICL, \citep{Delucia2007,Puchwein2010,Cui2014}. 
This is a diffuse faint component that grows over time with the mass assembly of the cluster, to which the relics of the interactions among galaxies (stellar streams and tidal tails) also contribute, that grows over time  with the mass assembly of the cluster \citep[see][as reviews]{Arnaboldi2010,Tutukov2011,Mihos2015}. 

From the theoretical side, semi-analytic models combined with cosmological N-body simulations have become very sophisticated, with detailed predictions about the structure of stellar halos, the ICL formation and the amount of substructures in various kinds of environment \citep[e.g.][and references therein]{Martel2012,Watson2012,Cooper2013,Pillepich2014,Contini2014,Cooper2015}.

Rich environments of galaxies, i.e., groups and clusters,  are therefore the appropriate sites to study the mass assembly processes that leads to the observed galaxy structure, stellar halos and ICL in order to test hierarchical formation theories at all scales.
In recent years, the advent of deep imaging surveys allows to study galaxy structures down to the faintest surface brightness levels and out to large distances from the galaxy centre  to map stellar halos and the ICL \citep{Mihos2005,Jan2010,MarDel2010,Roediger2011,Ferrarese2012,Duc2015,Dokkum2014,Munoz2015,Trujillo2015,Cap2015,Mihos2016,Merritt2016,Iodice2016,Crnojevic2016,Spavone2017}. 

The {\it Fornax Deep Survey (FDS)} is just a part of this campaign. FDS is a multiband ($u$,$g$,$r$ and $i$) imaging survey at the VLT Survey Telescope (VST) that aims to cover the whole Fornax cluster out to the virial radius \citep[$\sim0.7$~Mpc][]{Drinkwater2001},  with an area of about $26$~square degrees around the central galaxy NGC~1399 in the cluster core.
The Fornax cluster is the second most massive galaxy concentration within 20 Mpc, after the Virgo cluster, and it is among the richest nearby target to study galaxy evolution and dynamical harassment in dense environments.
Recent works indicate that it has a complex structure and the mass assembly is still ongoing.
The core is in a quite evolved phase \citep{Grillmair1994,Jordan2007}, since most of the bright ($m_B<15$~mag) cluster members are early-type galaxies \citep{Ferguson1989}, more than in the Virgo cluster.
It hosts a vast population of dwarf galaxies and ultra compact galaxies \citep{Munoz2015,Hilker2015a,Schulz2016}, an intra-cluster population of GCs \citep{Schuberth2010,Dabrusco2016} and of planetary nebulae \citep{Napolitano2003,McNeil2012}.
Dynamics on the large scale shows that a poor subgroup of galaxies on the SW, is falling into the cluster core along a cosmic web filament \citep{Drinkwater2001,Scharf2005}. The FDS also includes such an in falling structure of the Fornax cluster.
Therefore, the FDS will provide an unprecedented view of structures of galaxies, ranging from giant early-type galaxies to small spheroidal galaxies, a full characterisation of small stellar systems around galaxies, as globular clusters (GCs) and ultra-compact dwarf galaxies (UCDs) and, on the cluster scale, will provide hints on the galaxy evolution as function of the environment.

First results have provided the mosaic of $3\times2$ square degrees around the central galaxy NGC~1399\footnote{See the ESO photo release at https://www.eso.org/public/news/eso1612/}. The deep imaging allows us to map  the surface brightness out to the distance of $\sim 200$~kpc ($R\sim6R_e$) from the centre of NGC~1399 and down to $\mu_g \simeq 31$ mag/arcsec$^2$  \citep{Iodice2016}. In the intracluster region, on the west side of NGC~1399 and towards NGC~1387, we have detected a faint ($\mu_g \sim 29-30$~mag~arcsec$^{-2}$) stellar bridge, about 5~arcmin long ($\sim 29$~kpc), which could be due to the ongoing interaction between the two galaxies, where the outer envelope of NGC1387 on its east side is stripped away.
By using about 3000 candidate GCs extracted from the VST $ugri$ images covering the central $\sim 8.4$ deg$^2$ of the cluster, \citet{Dabrusco2016} traced the spatial distribution of candidate GCs in a region $\sim0.5$~deg$^2$ within the core of the Fornax cluster of galaxies. In particular, they confirmed the bridge between NGC~1399 and NGC~1387, finding an over density of blue GCs in this region, as previously suggested by \citet{Bassino2006}.

The FDS observations are planned to be completed at the end of 2017. A full description of the data will be presented soon in an forthcoming paper. In this current work, we focus on a new mosaic of the SW group of the Fornax cluster, in the $g$ and $r$ bands, which covers an area of  $3 \times 2$~square degrees around the central galaxy NGC~1316 (see Fig.~\ref{mosaic}).
In the following, we adopt a distance for NGC~1316 of $D=20.8\pm0.5$~Mpc \citep{Cantiello2013}, which yields an image scale of 101~parsecs/arcsec.
}

 \section{A literature review of NGC~1316}\label{n1316}

NGC~1316, also known as Fornax~A, is one of the most fascinating giant galaxies in the local universe. It is the brightest member of the Fornax cluster located in the in the subgroup to the SW of the cluster \citep{Drinkwater2001}, at about 4 degrees ($\sim1.5$~Mpc)  from NGC~1399.
{  Over the past years, NGC~1316 has been widely studied in a large wavelength range. The results are summarised in this section. The main observed properties show a
very complex structure for this galaxy and point towards a rich history of interaction events. }

\begin{figure*}[t]
\centering
\includegraphics[scale=0.8]{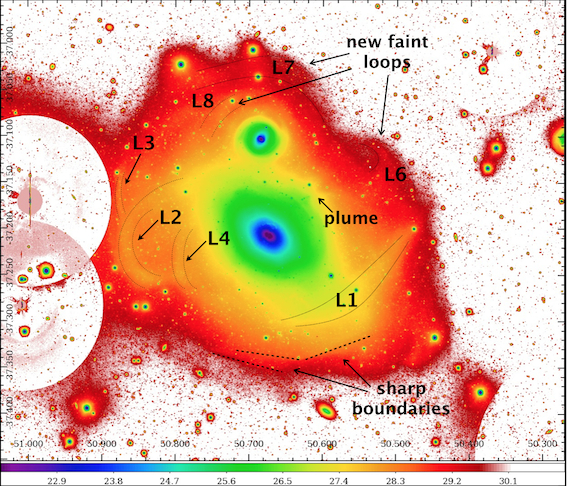}
\caption{\label{zoomcenter}  Structure of NGC~1316 as it appears in the $g$ band, plotted in surface brightness levels (shown in the colorbar). The image size is $\sim 1 \times 0.6$~square degrees. The most luminous substructure are marked on the image.
The known loops presented by \citet{Schweizer80} and \citet{Richtler2014} are L1, L2, L3 and L4, while the new and faintest ones found in the deep VST data are on the N-NW. The peculiar sharp boundaries on the S-SE side are delimited by the dashed lines. }
\end{figure*}

{  
{\it Morphology and structure in the optical and infrared -} 
NGC~1316 resembles an elongated spheroid characterised by pronounced and extended dust patches, more concentrated in the galaxy centre and best mapped by Hubble Space Telescope images \citep{Carlqvist2010}, and several ripples and loops, with different shape and luminosity (see Fig.~\ref{zoomcenter}). 
The first detailed description of all the substructures in NGC~1316 was given by \citet{Schweizer80}, based on wide-field photographic plates. In particular. he pointed out the giant  L5 loop on the SW regions of the galaxy outskirt and an elongated feature dubbed "plume" inside the galaxy, on the NW  (see Fig.~\ref{zoomcenter}).
Besides, \citet{Schweizer80} detected an inner rotating disk of ionised gas, along the minor axis of the galaxy, and a giant HII-region south of the centre. In a region where there is no evident sign of ongoing star formation, at a distance of 6 arcmin from the centre, 
\citet{MF1998} found an extended ($\sim1.5$ arcmin) emission line region (EELR). 
In the nucleus of the galaxy there is a kinematically decoupled cold stellar disk and a spiral-like structure of gas at about $5 -15$ arcsec from the centre \citep{Beletsky2011}.
The {\it Spitzer} images (at 8.0 $\mu$m) map the morphology of the dust emission in the inner 3~arcmin, which display a bar-like structure along the galaxy minor axis ending with a spiral arms, most pronounced on the NW side  \citep[see Fig.~3 of][]{Lanz2010}. 
\citet{Duah2016} suggested that the dust in this galaxy has an external origin, coming from a recent infall and disruption of one or more smaller gas-rich galaxies. From the {\it Spitzer} images, \citet{Lanz2010} estimated that the merger late-type galaxy had a stellar mass of $\sim 10^{10} M_{\odot}$ and an amount of gas of  $\sim 10^{9} M_{\odot}$.


{\it Emission from NGC1316 -}
The amount of gas in NGC~1316 is quite small. \citet{Horellou2001} found  $\sim 5\times10^8$~M$_{\odot}$ of molecular gas inside 2~arcmin, which is mainly associated to dust. The galaxy is devoid of neutral atomic gas to a limiting mass of $\sim10^8$~M$_{\odot}$, while some emission is detected in four clumps in the galaxy outskirts. Two of these coincide with the HII region and the EELR in the South \citep{Schweizer80,MF1998,Richtler2012} and  two others are on the NW side. The total gas-to-stellar mass ratio in NGC~1316 is $\sim10^{-4}$.

The X-ray emission of NGC~1316 was mapped with most of the X-ray satellites \citep{Fabbiano1992,Feigelson1995,Kim1998,Iyomoto1998,Kim2003,Konami2010}. As shown by \citet[][see Fig.6 of that paper]{MF1998}, the main contributions to the X-ray emission are with the galaxy centre and with two of the most luminous substructures in the light, dubbed by \citet{Schweizer80} as L1 in the South and L2 on East of the galaxy (see Fig.~\ref{zoomcenter}). There is an elongated region of  X-ray emission to the NW of NGC~1316, corresponding to the region of the plume.

In the radio domain, NGC~1316 is the third brightest nearby galaxy ($L_{radio} =259$~Jy), after NGC5128 (Centaurus~A) and M87 \citep{Ekers1983}, within 20~Mpc.
At a position angle of 110 degrees, 30 degrees away from the galaxy rotation axis, there are prominent radio lobes, spanning 33 arcmin, made by polarised filaments \citep{Fomalont1989}. In addition to the radio lobes, there is also an S-shaped nuclear radio jet \citep{Geldzahler1984}. A low-level emission bridge is detected between the two lobes, but it is displaced of about 8 arcmin from the centre of the galaxy. The authors claim that the offset bridge, which is unusual in radio galaxy, could be the remnant of the original bridge, formed with the lobes and aligned with them but disrupted by an in falling galaxy. From  combined Spitzer, Chandra, XMM-Netwton and VLA observations for NGC~1316 \citet{Lanz2010} gave a detailed description and formation of the radio lobes and X-ray cavities. 

{\it Globular cluster system -}
The globular clusters (GC) of NGC~1316 have been studied in detail over the past ten years, suggesting the presence of a complex GC system in terms of age and metallicities \citep[][and references therein]{Goudfrooij2012,Richtler2012}. Recent results by \citet{Sesto2016} indicate that the GCs system in NGC~1316 is made by four stellar cluster populations with different age, abundances and spatial distribution. Two of them have similar characteristics to the GCs observed in early-type galaxies. The dominant GC component has an intermediate-age stellar population ($\sim 5$ Gyr), and the younger GC population is about 1 Gyr. 
The presence of an intermediate-age stellar population was also pointed out by \citet{Cantiello2013} by studying the surface brightness fluctuations in NGC~1316.

{\it Stellar kinematics and dynamics -}
The stellar kinematics in NGC~1316 indicate that the rotation velocity along the major axis increases to $~150-170$~km/s at $R\sim90$~arcsec, and  decreases to $~100$~km/s at larger radii, out to $R\sim130$~arcsec. The velocity dispersion is quite high, being $~250$~km/s in the centre and decreasing to $~150$~km/s at $R\sim380$~arcsec \citep{Bedregal2006,Arnaboldi1998}.
The amount of rotation derived from the planetary nebulae (PNe) kinematics remains almost constant at about 100 km/s out to 500 arcsec from the centre \citep{McNeil2012}. 
The dynamical models based on both PNe and GCs indicate a high dark matter content, similar to what is found in giant ellipticals \citep{McNeil2012,Richtler2014} .}

\section{Observations and Data Reduction}\label{data}


The observations presented in this work are part of the ongoing {\it Fornax Deep Survey} (FDS) with the ESO VLT Survey
Telescope (VST) \citep[see][]{Iodice2016}. FDS is based on two Guaranteed Time Observation surveys, {\it FOCUS}\footnote{www.astro.rug.nl/fds}
(P.I. R. Peletier) and {\it VEGAS}\footnote{http://www.na.astro.it/vegas/VEGAS/Welcome.html} \citep[P.I. E. Iodice,][]{Cap2015}.

VST is a 2.6-m wide field optical survey telescope, located at Cerro Paranal in Chile \citep{Schipani2012}, 
equipped with the wide field camera OmegaCam, covering the optical wavelength range from 0.3
to 1.0 micron \citep{Kui2011}. The field of view of OmegaCam is $1 \times
1$~degree$^2$ and the mean pixel scale is 0.21~arcsec/pixel.

The observations were collected during a visitor mode run from 9 to 14 of November 2015 (runID: 096.B-0582(A)). 
Images were taken in  $g$ and $r$  bands\footnote{See at the following link for the filter description https://www.eso.org/sci/facilities/paranal/instruments/omegacam/inst.html} during dark time, in photometric conditions, with an average seeing between 0.6 and 1.1~arcsec. We covered an area of 6~deg$^2$ around NGC~1316.

{  The data were processed with the {\it VST-Tube} imaging
pipeline, whose main steps (pre-reduction, astrometric and photometric calibration, mosaic production) are discussed in details in  \citet{Grado2012} and in Appendix~A of \citet{Cap2015}. }

{   We used the {\it step-dither} observing strategy, already adopted  for the observations of the core of the Fornax cluster \citep{Iodice2016}. It consists of a cycle of short exposures (150 seconds) for each of the six fields acquired in the area of Fornax~A. For each field, we obtained 44 exposures in the $g$ band and 33 in the $r$ bands, giving a total exposure time of 1.83 hrs and 1.37 hrs in the $g$ and $r$ bands, respectively.
This observing strategy allows a very accurate estimate of the sky background around bright and extended galaxies.
For the FDS data set presented in this paper, the five (almost empty) fields around that containing NGC~1316 are used as sky frames. 
Two of them are on the East, two others on the West and one more on the North of NGC~1316.
For each night,  we derived an average sky frame using these 5 fields (where all bright sources are masked). 
The average sky frame is scaled and subtracted off from each science frame. 
In the first paper of the FDS,  \citet{Iodice2016} discussed in detail the possible contributions to the sky brightness (as galactic cirrus, zodiacal light and of the terrestrial airglow) and how to take them into account. The small contribution of the  smooth components listed above plus the extragalactic background light to the sky brightness is taken into account by the average sky frame derived for each observing night. In the sky-subtracted science frame only a possible differential component could remain,  which contributes to the ''residual fluctuations''\footnote{The residual fluctuations in the sky-subtracted images are the deviations from the the background in the science frame with respect to the average sky frame obtained by the  empty fields close to the target.} in the background and sets the accuracy of the sky-subtraction step.

In Fig.~\ref{mosaic} we show the sky-subtracted VST mosaic in the $g$  band of the central $3.6 \times 2$~degrees$^2$  around  NGC~1316. 
For all the Fornax cluster members in this area, included in the Fornax Cluster Catalogue by \citet{Ferguson1989},
the integrated magnitudes in the $g$ band and the $g-r$ integrated color inside the circular aperture at the radius corresponding to $\mu_g = 30$~mag/arcsec$^2$ are listed in Tab.~\ref{mag_galaxies}.}

The inner structure of NGC~1316 and the faint features in the outer envelope are presented and discussed in the next sections. 
The surface photometry of all the bright galaxies in this subgroup of the Fornax cluster is the subject of a forthcoming paper.


\begin{table*}
\caption{\label{mag_galaxies} Integrated magnitudes in the $g$ and the $g-r$ color for the member galaxies in the region of the Fornax cluster shown in Fig.~\ref{mosaic}.}
\begin{tabular}{lccccccc}
\hline\hline
object & $\alpha$ & $\delta$ & Morph Type & radial velocity & $r_{30}$ & $m_g$ & $g-r$\\
           & h:m:s     & d:m:s     &                     & km/s & [arcsec]           & [mag] & [mag]\\
(1) & (2) & (3) & (4) & (5) & (6) & (7) & (8) \\
\hline
\hline
NGC~1310 & 03 21 03.4 & -37 06 06& SB & 1805 &105 & $12.351\pm0.004$ & $0.544\pm0.006$\\
NGC~1316 & 03 22 41.7 & -37 12 30 & SAB0 & 1760 & 1985 & $8.43\pm0.003$ & $0.717\pm0.005$\\
NGC~1316C & 03 24 58.4 & -37 00 34 & SA & 1800 & 66.15 & $13.720\pm0.006$ & $0.51\pm0.01$\\
NGC~1317 & 03 22 44.3 & -37 06 13 & SAB & 1941 & 115.5 & $11.012\pm0.003$ & $0.729\pm0.004$\\
NGC~1326 & 03 23 56.4  & -36 27 53 & SB & 1360 & 178.5 & $10.491\pm0.003$ & $0.646\pm0.004$\\
NGC~1326A & 03 25 08.5 & -36 21 50 & SB & 1831 & 63 & $13.697\pm0.004$ & $0.378\pm0.006$\\
NGC~1326B & 03 25 20.3 & -36 23 06 & SB & 999 & 118.65 & $12.584\pm0.006$ & $0.29\pm0.01$\\
NGC~1341 & 03 27 58.4 & -37 09 00 & SAB & 1876 & 82.95 & $11.918\pm0.003$ & $0.399\pm0.004$\\
FCC019 & 03 22 22.7 & -37 23 54  & dS0 & 1407 & 39.06 & $15.119\pm0.007$ & $0.58\pm0.01$\\
FCC023 & 03 23 24.7 & -36 53 11  & ImV or dE5 & ... & 10.5 & $19.03\pm0.02$ & $0.60\pm0.03$\\
FCC025 & 03 23 33.5 & -36 58 52  & dE0 & .... & 16.8 & $17.66\pm0.01$ & $0.54\pm0.02$\\
FCC028 & 03 23 54.2 & -37 30 33 & SB & .... & 10.5 & $13.357\pm0.004$ & $0.536\pm0.006$\\
FCC035 & 03 25 04.2 & -36 55 39  & SmIV & 1800 & 36.75 & $15.028\pm0.006$ & $0.23\pm0.01$\\
FCC038 & 03 25 09.4 & -36 33 07  & dS0 & .... & 17.85 & $17.47\pm0.01$ & $0.49\pm0.02$\\
FCC040 & 03 25 24.8 & -36 44 36  & dE4 & .... & 16.8 & $17.70\pm0.01$ & $0.56\pm0.02$\\
FCC046 & 03 26 25.0 & -37 07 34  & dE4 & .... & 37.8 & $15.354\pm0.008$ & $0.46\pm0.01$\\
FCC049 & 03 26 43.5 & -37 17 51  & dE4 & .... & 13.86 & $19.15\pm0.03$ & $0.47\pm0.05$\\
FCC051 & 03 26 53.1 & -36 47 51  & dE4 & .... & 9.87 & $18.04\pm0.01$ & $0.33\pm0.02$\\
%
%
FCC058 & 03 27 34.6 & -37 30 11  & dE4 & .... & 17.75 & $17.79\pm0.01$ & $0.76\pm0.02$\\
FCC060 & 03 27 54.0 & -37 22 27  & ImV or dE2 & 1407 & 18.48 & $18.31\pm0.02$ & $0.35\pm0.04$\\
\hline
\end{tabular}
\tablecomments{{\it Col.1 -} Fornax cluster members from \citet{Ferguson1989}. {\it Col.2 and Col.3 -} Right ascension and declination.  {\it Col.4 and Col.5 -} Morphological type and Heliocentric radial velocity given by NED.  {\it Col.6 -}aperture radius in arcsec corresponding to the surface brightness $\mu_g=30$~mag/arcsec$^2$. {\it Col.7 -} integrated  magnitude in the $g$ band, derived inside a circular aperture with $r=r_{30}$. Values were corrected for the Galactic extinction by using the absorption coefficient derived according to \citet{Schlegel98}. {\it Col.8 -} integrated $g-r$ color.}
\end{table*}


\section{The complex structure of NGC~1316}\label{phot}

The structure of NGC~1316 has been widely described in the literature \citep{Schweizer80,MF1998,Richtler2012,Richtler2014}. To date, the VST mosaics in the $g$ (see Fig.~\ref{mosaic}) and $r$ bands cover the most extended area observed around NGC~1316. 
{  In this section we describe the morphology and colour distribution of NGC~1316, focusing on the new main features revealed by the deep VST images.}

\subsection{Morphology: the new faint features in the galaxy outskirts}\label{outskirt}

\begin{figure*}[t]
\centering
\includegraphics[scale=0.7]{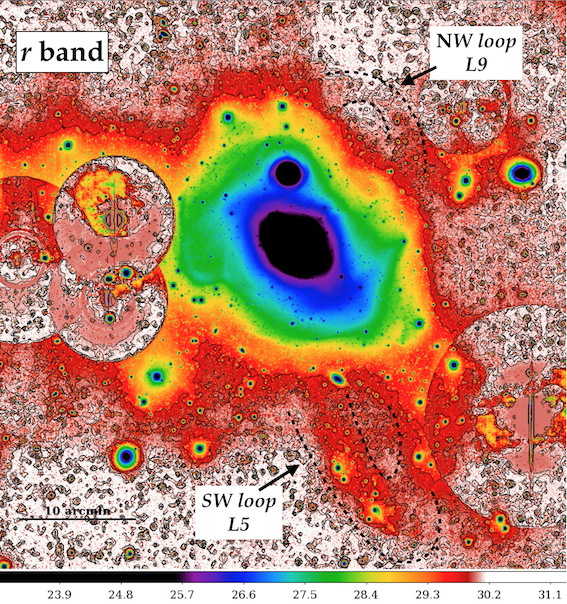}
\caption{\label{n1316mur}  Enlarged regions of the mosaic shown in Fig.~\ref{mosaic}, of about $1\times1$~square degrees around NGC~1316 in $r$ band, plotted in surface brightness levels (shown in the colorbar). Black lines are the following surface brightness levels:  29.65, 29.9, 30.05, 30.1 mag/arcsec$^2$. 
The brightest part of close stars were subtracted off. The extended loop in the SW area (L5) and the new faint loop on the NW (L9) are marked on the image. Right Ascension and Declination in degrees are reported on the horizontal and vertical axes.}
\end{figure*}

{  In Fig.~\ref{zoomcenter} we show a smaller region of the $g$-band mosaic (plotted in surface brightness levels, where the brightest stars are modelled and subtracted), zoomed on the galaxy centre.   There is a wealth of  loops that characterise the outskirts of NGC~1316, down to surface brightness down to $\mu_g \sim 30$~mag/arcsec$^2$.
Shells and dusty features perturb the inner and luminous regions of the galaxy.
We have marked the known loops (L1, L2, L3, L4)  and substructures (sharp boundaries and the plume)  identified and discussed by \citet{Schweizer80} and  \citet{Richtler2012}. In addition to these, the deep VST images allow us to detect at least three more  faint loops on the N-NW side of the envelope, at the level of  $\mu_g \sim 29 - 30$~mag/arcsec$^2$, which we named  L6, L7 and L8. The loop L6 in the NW, appears as a very curved arc emerging from the bright west edge of L1.
The large area covered by the VST mosaic and the large integration time allow us to confirm the presence of the giant loop L5 identified by \citet{Schweizer80} on the photographic plates, in the SW region (see Fig.~\ref{n1316mur}). It extends out to about 0.44~degree from the galaxy centre, corresponding to a projected distance of about 160~kpc. In addition, we notice a very faint and less extended (15~arcmin $\sim91$~kpc) loop, protruding from the outer envelope on the NW side of the galaxy, identified as L9 (see Fig.~\ref{n1316mur}).  The VST images allow us to point out the structure of the SW giant loop L5: it has a  filamentary appearance close to the galaxy, while  several bright knots dominate the emission on its SE part. On the West side it seems to have a more diffuse structure, but here the light is contaminated by the faint halo of the close bright star. The brightest knots and filaments have an $r$-band surface magnitudes in the range $29 - 30$~mag/arcsec$^2$ (see Fig.~\ref{n1316mur}). The NW small loop L9 is even fainter than this. }


\subsection{The  colour distribution}\label{color}

\begin{figure*}[t]
\centering
\includegraphics[scale=0.8]{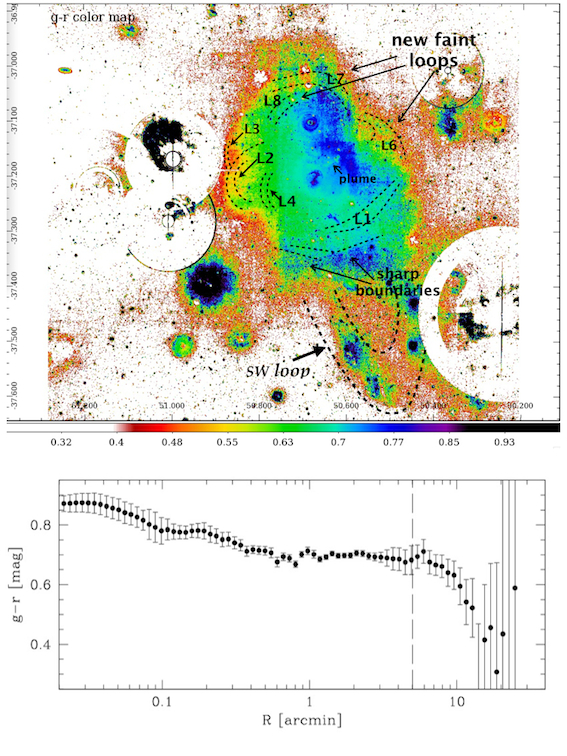}
\caption{\label{colormap}  {\it Top panel -} $g-r$ colormap in the same area shown in Fig.~\ref{n1316mur}, where the most luminous substructure are marked on the image. The $g-r$ levels are shown in the colorbar. {\it Bottom panel -} Azimuthally-averaged $g-r$ color profiles for NGC~1316  as function of the fitted ellipses semi-major axis. The vertical  long-dashed line indicate the transition radius at $R = 5.5$~arcmin, between the main galaxy body in NGC~1316, fitted by a Sersic $n\sim4$ law, and its outer exponential stellar halo (see Sec.~\ref{fit} for details).}
\end{figure*}

In Fig.~\ref{colormap} (top panel) we show the $g-r$ colormap, where we marked the regions of all features identified in NGC~1316 and described in the previous sections.

\begin{figure}[t]
\centering
\includegraphics[scale=0.35]{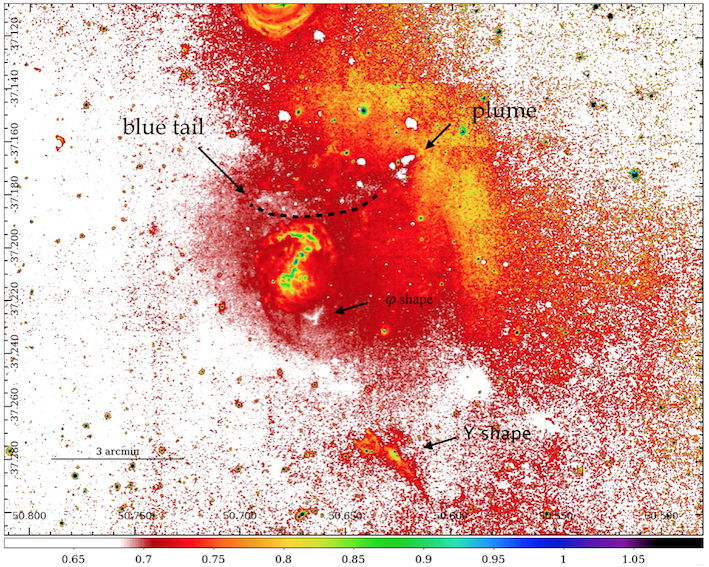}
\caption{\label{colormap_zoom}  Zoom of the $g-r$ colour map in galaxy centre from the image shown in Fig.~\ref{colormap}. North is up and East is on the left. }
\end{figure}

The $g-r$ map points out two main characteristics in the colour distribution: {\it i)} the color map is asymmetric, being redder on the NW side ($0.8 \le g-r \le 1$~mag) than the corresponding one on the SE side ($0.6 \le g-r \le 0.8$~mag). This was already noticed by \citet{Richtler2014}, but on a smaller area close to the centre; {\it ii)}  in correspondence of the sharp boundaries there is a ``sharp" change also in the color, where  south parts are redder.
On the North, the colour map shows that stars and dust in the close companion NGC~1317 follow an unperturbed spiral structure and the disk appears intact. As already stressed by  \citet{Richtler2014}, this suggests that, even if the two galaxies appear very close in projection, an interaction between them can be reasonably excluded.
{  The colour distribution of  the giant SW loop L5 is mapped for the first time.  Differently from  what observed for the outermost loops (L2, L6, L7 and L8), L5 is redder, with average color similar to the inner region of the galaxy. There are two extended areas of L5, close to the bright knots identified in the light distribution that have  $0.69\le g-r \le 0.78$~mag.

The centre of the galaxy is very red, with $g-r\sim0.8$~mag. For $R\ge3$~arcmin, the average colour is $g-r\sim0.65$~mag and it is bluer, down to 0.5~mag, in the galaxy outskirts. The outermost loops (L2  and the new ones L6, L7 and L8) have bluer colors $0.5\le g-r \le0.65$~mag than most of the galaxy bulk. 

In the inner 3~arcmin, the color map reveals the spiral-like system of dust lanes, more elongated towards N-NE (see Fig.~\ref{colormap_zoom}). We catch  two other "dust arms" that are more extended but less dense than the nuclear counterparts. 
On the NW region of the galaxy there are several blue patches  with $g-r \le 0.6-0.7$~mag.  One of the most extended is the {\it plume}, as named by \citet{Schweizer80}, which is about 2~arcmin ($\sim12$~kpc) long on the NW side. It has a very elongated shape, with the bluest colors concentrated in knots on the NW edge ($g-r \sim 0.6$~mag), while it appears in a more diffuse shape toward SE.  Close to the plume, on the north with respect to the galaxy centre, we detect a blue tail, with similar colors to the plume (see Fig.~\ref{colormap_zoom}). Its shape and colors might suggest a possible connection between the two features.


On the south side of the nucleus, there are two peculiar features: the "$\phi$-shape", firstly detected by \citet{Richtler2012}, which is bluer ($g-r \sim 0.68$~mag) than the background galaxy light, and the "Y-shape" located at about 4.6~arcmin SW from the galaxy centre, with redder colours ($0.8 \le g-r \le 1$~mag) than the average adjacent regions, so it is a probably patchy dust. }

 

\section{Isophote analysis: light distribution of NGC~1316 out to 200 kpc}\label{ellipse}

\begin{figure*}[t]
\includegraphics[scale=0.46]{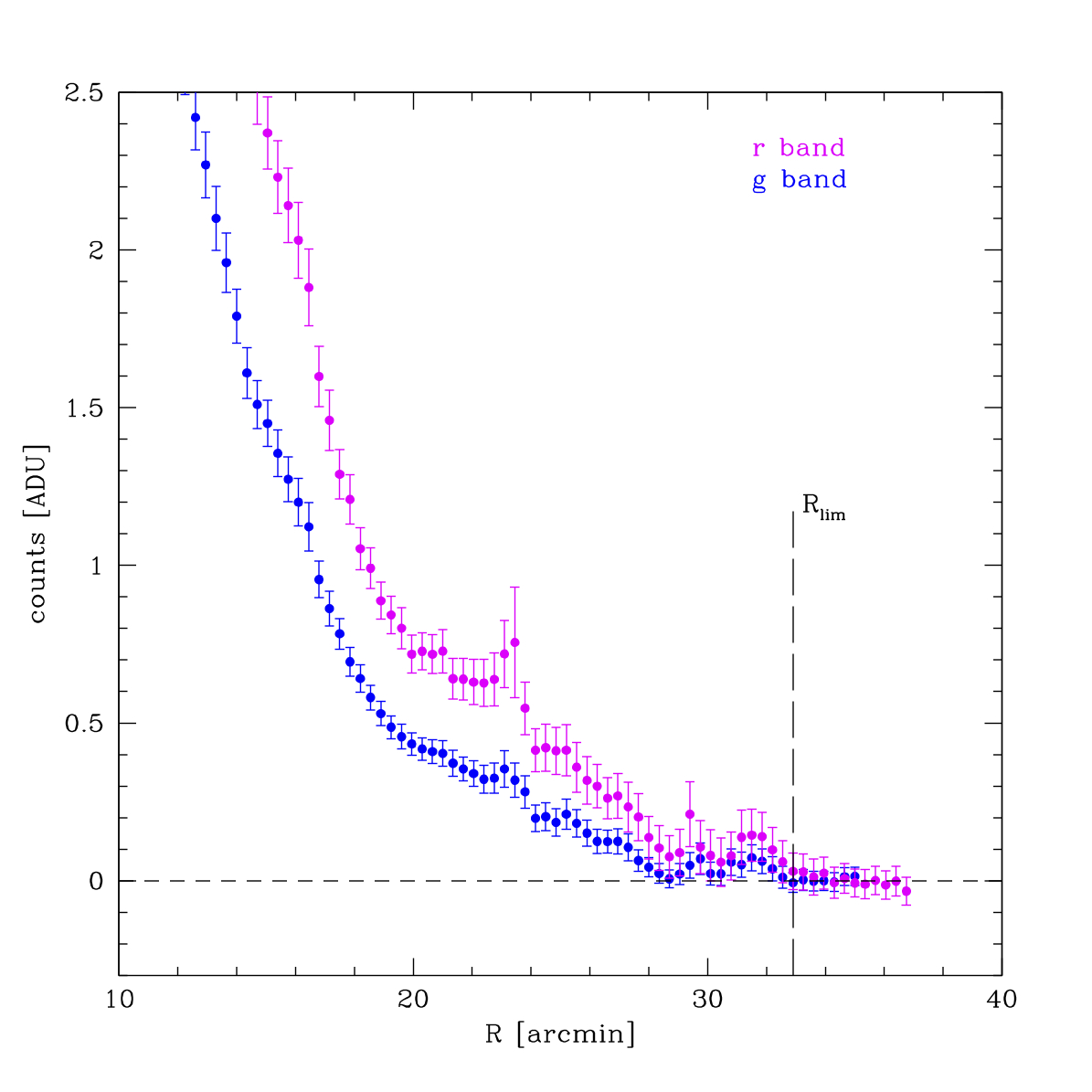}
\includegraphics[scale=0.44]{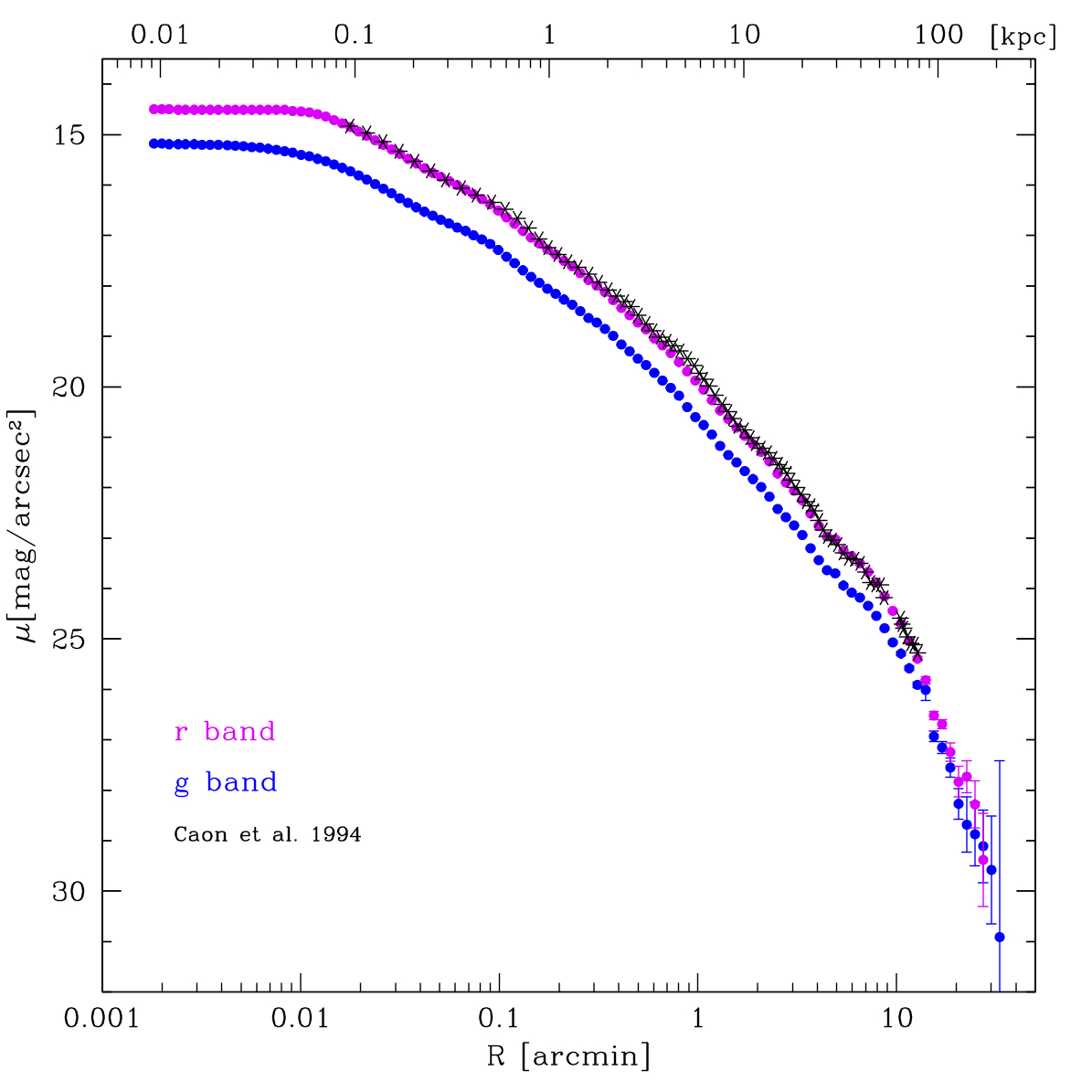}
\caption{\label{prof_sky} {\it Left panel} - Intensity profiles of NGC~1316 in the $g$ (blue points) and $r$ (magenta points) bands out to the regions of the background level (short-dashed black line). The outer radius from the galaxy centre where the galaxy's light blends into the background level is at $R_{lim} \simeq 33$~arcmin (long dashed line). 
{\it Right panel} - azimuthally-averaged surface brightness profiles of NGC~1316 in the $g$ and $r$ bands derived from VST mosaics,  compared with literature data from \citet{Caon1994} (black asterisks), which is obtained by averaging the light profiles obtained along the major and minor axes of the galaxy.}
\end{figure*}

\begin{figure}
\includegraphics[scale=0.42]{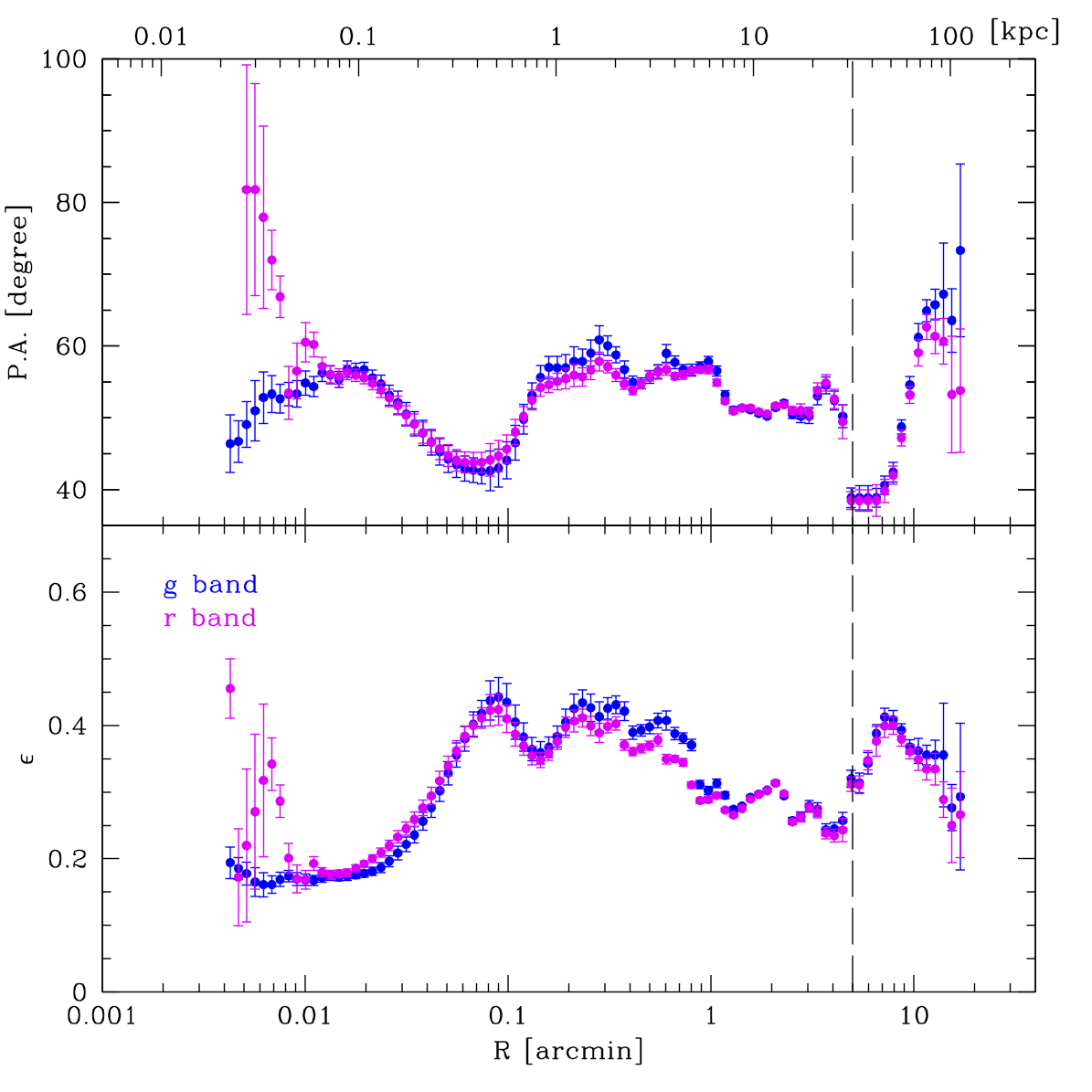}
\caption{\label{epsPA} Ellipticity (bottom panel) and P.A. (top panel) profiles for NGC~1316 derived by fitting the isophotes in the $g$ (blue points) and $r$ (magenta points) band images. The vertical long-dashed line indicate the transition radii at $R = 5$~arcmin, between the main galaxy body, fitted by a Sersic $n\sim4$ law, and the outer stellar exponential halo (see Sec.~\ref{fit} for details).}
\end{figure}

On the full sky-subtracted mosaic, in each band, we extracted the azimuthally-averaged intensity profile by using the IRAF task ELLIPSE. All the bright sources (galaxies and stars) and background objects have been accurately masked.
The fit is performed in elliptical annuli centred on NGC~1316 and the semi-major axis extends  out to the edge of the mosaic.
We derived  the azimuthally averaged surface brightness profiles (see Fig.~\ref{prof_sky}), the position angle (P.A.) and ellipticity ($\epsilon = 1 - b/a$, where $b/a$ is the axial ratio of the ellipses) profiles (see Fig.~\ref{epsPA}), and the colour $g-r$ profile (see Fig.~\ref{colormap}, bottom panel) as a function of the semi-major axis.
From the intensity profiles in both $g$ and $r$ bands, shown in the left panel of Fig.~\ref{prof_sky}, we estimated the outermost radius $R_{lim}$ from the centre of the galaxy where the galaxy's light blends into the average background level, which is the residual 
after subtracting the sky frame, and therefore very close to zero \citep[see][]{Iodice2016}. 
For both the $g$ and $r$ bands, we derived $R_{lim} \simeq 33$~arcmin ($\sim 200$~kpc). 
At this distance from the galaxy centre, the surface brightness profiles reach to  $\mu_g = 30 \pm 1$~mag arcsec$^{-2}$ and
$\mu_r = 29 \pm 1$~mag arcsec$^{-2}$ (see Fig.~\ref{prof_sky}, right panel).
The error estimates on the surface brightness magnitudes take the uncertainties  on the photometric calibration ($\sim 0.001 - 0.002$~mag) and sky subtraction ($\sim 0.02 - 0.03$~counts) into account \citep{Cap2015,Iodice2016}.
%
{  The light profiles in $g$ and $r$ bands are three times more extended and about five times deeper than the light profile published by  \citet{Caon1994} in the B band\footnote{Following the colour  transformation by \citet{Fukugita1996} $B-r=1.33$, the B band profile is shifted to the $r$} (see right panel of Fig.~\ref{prof_sky}). Besides, the agreement between the surface brightness derived from VST image in the $r$ band and literature data is satisfactory at all radii.}

{   Using  the growth curve,  we derived the total magnitudes and effective radii ($R_e$) inside 33~arcmin in the $g$ and $r$ bands. In the $g$ band, $m_{tot}=8.43\pm0.003$~mag and the $r$ band  $m_{tot}=7.713\pm0.002$~mag (see also Tab.~\ref{mag_galaxies}). The absolute magnitudes and the total luminosity in solar units\footnote{We used  $m_g=5.45$~mag and $m_r=4.76$~mag for the solar magnitudes in $g$ and $r$ bands, respectively, given at the following site http://www.ucolick.org/~cnaw/sun.html} in the $g$ band are $M_g=-23.16$~mag and $L=3.11 \times 10^{11} L_{\odot}$.
The effective radii in the $g$ and $r$ bands are $R_e = 2.22$~arcmin ($\sim13$~kpc) and $R_e = 2.23$~arcmin, respectively.}
Therefore, the surface brightness profiles extend out to $\sim15R_e$.
The  average $g-r \sim 0.72$~mag,  and, inside $R_e$, it is $g - r \sim 0.74$~mag.

{  Taking into account the average $g-r$ color and the age estimate of 2-3 Gyrs, given by the color distribution of the GCs \citep{Richtler2012}, by using the stellar population synthesis model \citep{Vazdekis2012,Ricciardelli2012} we estimate a stellar mass-to-light ratio (M/L) in the $g$ band in the range $1.67 \le M/L \le 2.66$. Therefore, the total stellar mass is in the range $5.2 \le M \le 8.3 \times 10^{11}$~M$_{\odot}$.}


{  The  P.A. and ellipticity profiles are shown in Fig.~\ref{epsPA}. For $R \ge 0.01$~arcmin, i.e. outside the central seeing  region, the P.A. and ellipticity profiles in the $g$ and $r$ bands are consistent. For $R \le 0.1$~arcmin, a strong twisting of about 40 degrees  and an increasing ellipticity from 0.2 to 0.4 are observed. For $0.1 \le R \le 4$~arcmin (0.6 - 24 kpc), the P.A. is almost constant at $\sim50 - 55$~degrees and ellipticity shows a decline from 0.4 to 0.25.
At larger radii, for $R\ge5$~arcmin, an abrupt variation of about 30 degrees is observed in P.A. while the ellipticity increases again to about 0.4. }



\subsection{Fit of the light profile}\label{fit}

\begin{figure*}[t]
\includegraphics[scale=0.45]{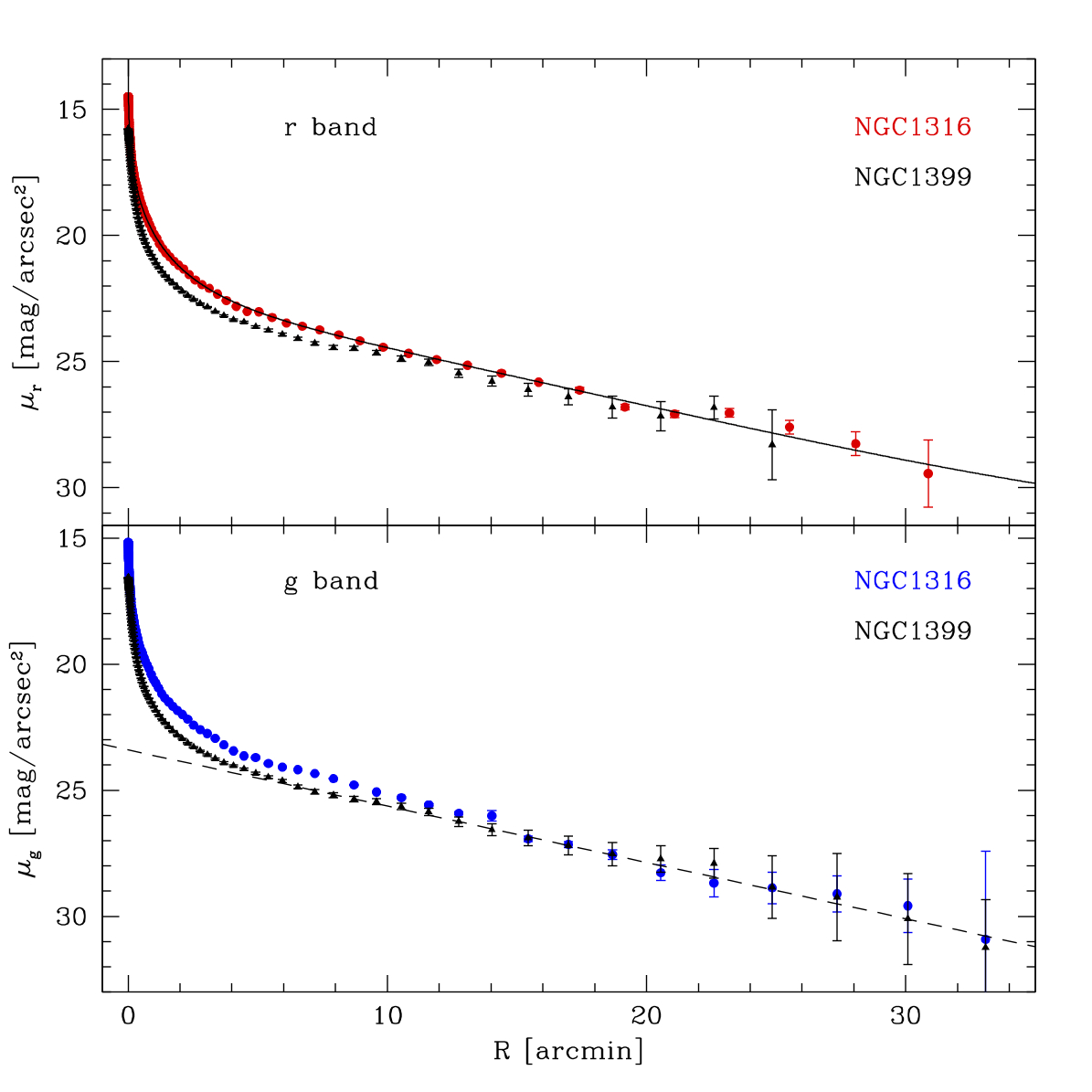}
\includegraphics[scale=0.45]{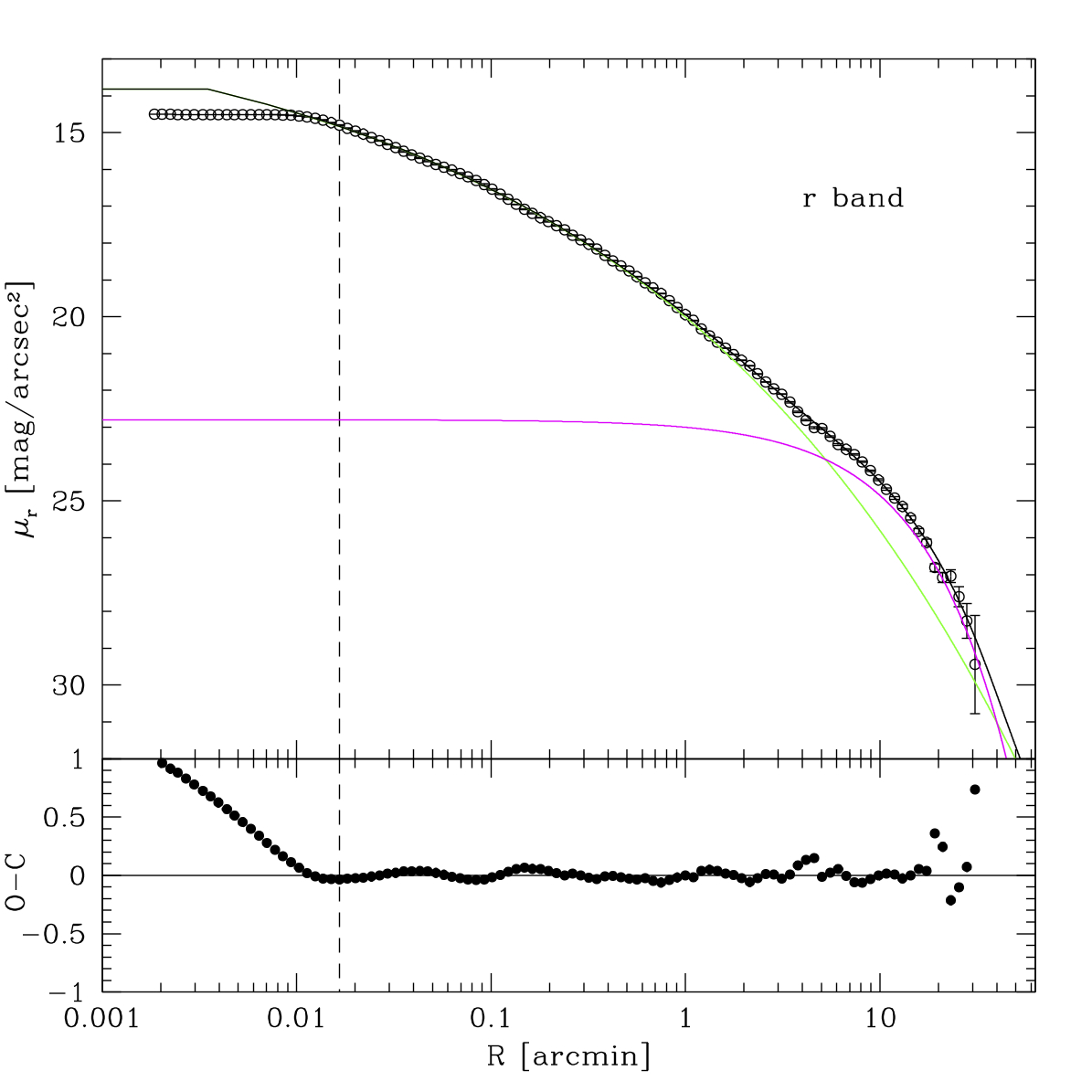}
\caption{\label{conf1399} {\it Left panel} - Azimuthally-averaged surface brightness profiles of NGC~1316 in the $g$ (bottom) and $r$ (top) bands derived from VST mosaic, in linear scale, compared with those derived for NGC~1399 (black asterisks) from \citet{Iodice2016}. The continuous line is  the resulting fit to the $r$-band profile. The dashed line is the exponential law adopted to fit the stellar halo in the $g$-band profile of NGC~1399 \citep[see][]{Iodice2016}.
{\it Right panel} - In the upper box there are the azimuthally-averaged surface brightness profile of NGC~1316 in the  $r$  band (open circles) and the resulting best fit obtained with a Sersic law (green line) and an exponential law (magenta line) to match the outer stellar halo (blue line). The black solid line is the sum of the three components. The vertical dashed line delimits the region affected by the seeing, which is excluded from the fit.
In the bottom box are plotted the residuals of the fit (observed minus fitted surface brightnesses).}
\end{figure*}

{  In order to define the scales of the main components dominating the light distribution in NGC~1316, we performed a multi-component fit to reproduce the surface brightness profile, as done for the brightest cluster galaxies (BCGs) at the centre of the groups and clusters \citep{Seigar2007,Donzelli2011,Arnaboldi2012,Cooper2013,Huang2013,Cooper2015b,Iodice2016,Rodriguez2016}. BCGs consist of at least two components, the bright spheroidal body and the outer and very extended,  moderately flattened, stellar halo.  

In Fig.~\ref{conf1399} (left panel) we compare the azimuthally-averaged surface brightness profiles of NGC~1316 with those derived for NGC~1399 \citep{Iodice2016}, in the $g$ and $r$ bands. Overall, the light profiles are very similar.
Inside $R\le10$~arcmin, NGC~1316 is brighter than NGC~1399.  At larger radii, the flux levels and the exponential decline of the surface brightness profiles of both galaxies are comparable. Given that, according to  \citet{Iodice2016}, we used the superposition of a Sersic law plus an exponential function to fit the light profile of NGC~1316\footnote{The Sersic law is given by the equation 
$\mu (R) = \mu_e + k(n) \left[ \left( \frac{R}{r_e}\right)^{1/n} -1\right]$
where $R$ is the galactocentric distance, $r_e$ and $\mu_e$ are the
effective radius and effective surface brightness respectively, $k(n)=2.17 n - 0.355$.
The exponential law is given by 
$\mu(R)= \mu_{0} + 1.086 \times R/r_{h}$
where, $\mu_{0}$ and $r_{h}$ are the central surface brightness and scale length \citep{Iodice2016}.}.

We performed a least-square weighted fit\footnote{To perform the least-square weighted fit we used the MINUIT package \citep{James1975}, which is a physics analysis tool for function minimization and error analysis  distributed by the CERN (see https://www.cern.ch/minuit). As user-defined function we adopted the {\it chisquare} defined as $
\chi^2 \left(\alpha \right) = \sum \frac{\left[ f(x_i,\alpha) - e_i \right]^2}{\sigma_i^2}$,
where $\alpha$ is the vector of free parameters being fitted, and the $\sigma_i$ are the uncertainties in the individual measurements $e_i$. We computed the reduced chi-square $\tilde{\chi^2} = \chi^2/N$, where N is the number of measured values. The best is given when $\tilde{\chi^2} \sim 1$.}
of the azimuthally averaged surface brightness profile in the $r$ band, which is less perturbed by dust absorption in the centre, for $R \geq 0.02$~arcmin, to exclude the central seeing-dominated regions.
The structural parameters and the $\tilde{\chi^2}$ values are listed in Tab.~\ref{param_fit}. 
Results are shown in the right panel of Fig.~\ref{conf1399}.
We found that inside $R\le5.5$~arcmin the Sersic law coincides with a de  Vaucouleurs  law, since $n\sim4$. This component accounts for the inner and most luminous part of the galaxy, the {\it central spheroid}.
For $R \ge 5.5$~arcmin, the light profile is well reproduced by the exponential function.
The total magnitude obtained from the fit is $\sim0.5$~mag brighter  than that measured from the growth curve. Such a difference is due to the exclusion of the central seeing-dominated regions ($R\leq0.02$~arcmin)  from the fit.

The above analysis allows to conclude that there are two main components dominating the light distribution in NGC~1316 and to estimate the {\it transition radius} between them at $R=5.5$~arcmin. The presence of the two components is also consistent with the variations measured in the isophotes' ellipticity ($\epsilon$) and P.A. profiles, in correspondence of $R=5.5$~arcmin, where for $R\ge5.5$~arcmin,  $\epsilon$ increases from 0.22 to 0.4, the P.A. shows a  twisting by about 30 degrees SE (see Fig.~\ref{epsPA}). The central spheroid is redder ($g-r \sim 0.7$~mag) than the envelope, which shows a gradients towards bluer colors from $g-r \sim 0.7$~mag at  $R=5.5$~arcmin down to $g-r \sim 0.4$~mag at  $R=20$~arcmin (see bottom panel of Fig.~\ref{colormap}).

As for NGC~1399, the outer exponential component represents physically the {\it stellar envelope}, which has a central surface brightness  $\mu_0=22.79 \pm 0.10$~mag/arcsec$^{2}$ and a scale length $r_h=317 \pm 17 $~arcsec ($\sim 32$~kpc) in the $r$ band. These values are comparable with those estimated for the stellar exponential envelope of NGC~1399  \citep{Iodice2016}, which are $\mu_0 = 22.6\pm0.1$~mag/arcsec$^{2}$ and $r_h=292 \pm 4 $~arcsec ($\sim 28$~kpc) in the $r$ band.
The fraction of the stellar envelope in NGC~1316 with respect to the total luminosity is $\sim0.36$, which is almost half of  the 60\% estimated in NGC~1399 \citep{Iodice2016}.  }

\begin{table*}
\caption{\label{param_fit} Structural parameters derived from the multi-component fit of the surface brightness profile in the $r$ band for NGC~1316.}
\begin{tabular}{ccccccccc}
\hline\hline
\smallskip
${\mu_e}$ & ${r_e}$ & $n$ &  ${\mu_0}$ & ${r_h}$ &  $m_{tot}^{ser}$ & $m_{tot}^{exp}$ & $m_{tot}$ & $\tilde{\chi}^2$ \\
mag/arcsec$^2$] & [arcsec] & & [mag/arcsec$^2$] & [arcsec]  & [mag] & [mag] &[mag] & \\
(1) & (2) & (3) & (4) & (5) & (6) & (7) & (8) & (9)  \\
&&&&&&&&\\
$20.76\pm0.05$ & $87\pm2$ & $4.49\pm0.05$ & $22.79\pm0.10$ & $317\pm17$ & 7.67 & 8.29 & 7.18 &  1.035 \\
\hline
\end{tabular}
\tablecomments{{\it Col.1 - Col.5} Structural parameters characterising the empirical laws adopted to fit the surface brightness profile. The effective surface brightness $\mu_e$ and the central surface brightness $\mu_0$ are in mag/arcsec$^2$. The effective radius $r_e$ and the scale length $r_h$ are in arcsec. $m_{tot}$ is the total magnitude of each component. {\it Col.6 - Col.8} Total magnitudes for each component (Col.6 \& Col.7) and total magnitude of the fitted profile (Col.8).
{\it Col.9} Reduced $\tilde{\chi^2}$. The confidence level of each least-square fit is about 50\%.}
\end{table*}


\section{Behind the light: loops, ripples and accretion remnants inside NGC~1316}\label{2Dmodel}

\begin{figure*}[t]
\centering
\includegraphics[scale=0.7]{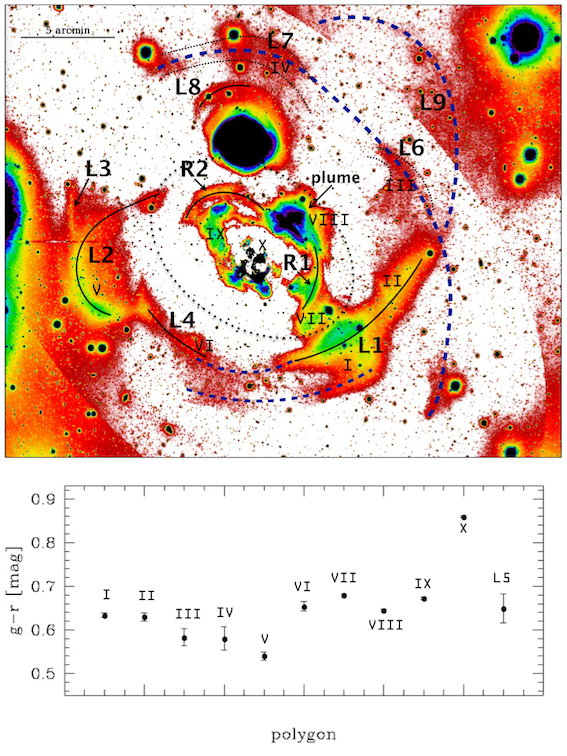}
\caption{\label{bmodel} {\it Top panel -} Residual map obtained by the ratio of the image and the 2-dimensional model in the $r$ band. The image size is $\sim 29.2 \times 23.3$~arcmin. The most luminous substructure are marked on the image.
The known loops presented by \citet{Schweizer80} and \citet{Richtler2014} are L1, L2, L3 and L4. The blu-dashed lines trace the  new and faintest features found in the deep VST data (see also Fig.~\ref{n1316mur}. Numbers from I to X indicates the regions where we computed the integrated $g-r$ colors shown in the bottom panel. The dotted ellipses correspond to the isophote at the transition radius $R = 5$~arcmin (see Sec.~\ref{fit} for details).
{\it Bottom panel -} Integrated $g-r$ colors in the 10 regions marked on the residual map (from I to X) and that computed for the SW loop L5.}
\end{figure*}

{  From the isophote fit performed in the  $r$ band (see Sec.~\ref{ellipse}), which is less perturbed by dust in the central regions than the g-band, we built a 2-dimensional (2D) model of the light distribution, by using the BMODEL task in IRAF. We derived the residual map as the ratio between the image of the galaxy and its 2D model. The result is shown in Fig.~\ref{bmodel} (top panel), where we marked the most luminous ripples and loops already described in Sec.~\ref{outskirt}.}

\begin{figure}
\centering
\includegraphics[scale=0.4]{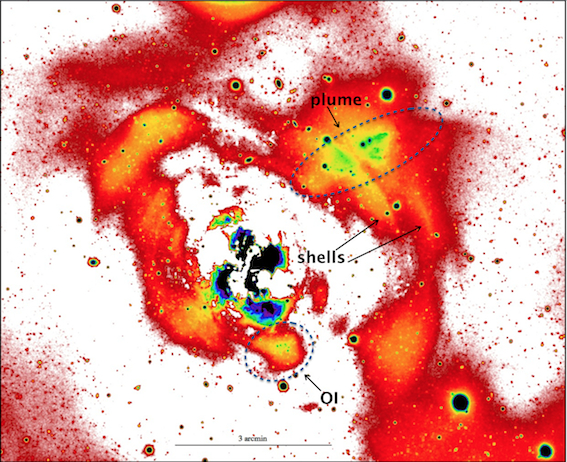}
\caption{\label{bmodel_zoom} Zoom into the central regions of the residual map shown in Fig.~\ref{bmodel}. The image size is $10.85\times8.9$~arcmin. The plume and the ``OI'' substructures are encircled by the blue dashed line.}
\end{figure}

{  {\it Morphology -} Blended into the galaxy's light,  in addition to the old and new substructures known for NGC~1316 (see  Sec.~\ref{outskirt}), the VST $r$-band residual image reveals {\it i)} a new and faint ``bridge'' connecting L1 with L4 and  
{\it ii)} on the West side, two shell-like features protruding from the south part of the plume (see Fig.~\ref{bmodel_zoom}). In projection, the inner shell is at about 2.7~arcmin from the centre, while the outer one is at  about 3.5~arcmin. They are very red, with average colors $g-r\sim0.8$~mag similar to those measured in the galaxy centre (see Fig.~\ref{colormap}). Moreover, 
the residual image brings to light the impressive coherent spiral-like pattern of ripples up to the centre of the galaxy and, on the SW at about 1.6~arcmin from the centre, the bright over-density region, labelled as ``O1'' by \citet{Richtler2014}. This is populated by several compact sources (see Fig.~\ref{bmodel_zoom}), which might be star clusters.  \citet{Richtler2014} suggested that, similarly to the plume, it could be another remnant of a dwarf galaxy.

The bright ripples R1 and R2, the plume and shells are features inside the transition radius $R=5.5$~arcmin (see Sec.~\ref{fit}), therefore they are part of the inner spheroidal component of the galaxy. All the loops from L1 to L9 contributes to the light of the outer stellar envelope.}


{  {\it Integrated colors -} We derived the integrated $g-r$ colors in several loops, ripples and substructures (i.e. the plume). The main aim is to compare them with the average value of the whole galaxy ($g-r = 0.717 \pm 0.005$~mag, see Tab.~\ref{mag_galaxies}): this analysis helps to discriminate their origin. 
Regions were selected on the residual map and polygons were defined covering the L1, L2, L3 and L4 (including the new faint bridge towards L1) loops, the plume region, ripples R1 and R2 (which includes the bright part close to the centre), the new two faint loops L6 and L7 on the  NW  of NGC~1316  (see details in Fig.~\ref{bmodel}) and the loop L5 in the south.
Results are shown in the bottom panel of Fig.~\ref{bmodel} and in Tab.~\ref{color_poly}. 

%

By analysing them we conclude that 
{\it i)} loops L1 and L4 seem to be a single structure, since they have similar colors ($g-r \simeq 0.63-0.65$~mag) and a faint bridge is detected between them (see the blue-dashed lines in Fig.~\ref{bmodel});
{\it ii)} the regions III and IV covering the new two faint loops L6 and L7 have similar colors ($g-r \simeq 0.58$~mag), thus they could be the brightest part of a single giant loop in the galaxy envelope, connecting also two others faint features on the south and on the north (see the blue-dashed lines in Fig.~\ref{bmodel}), with a diameter  $\ge 10$~arcmin; 
{\it iii)} loops L2 and L3, being about 0.13~mag bluer than the average color of NGC~1316, could have an external origin, while the giant SW loop L5 that has comparable colours to those of the inner regions could be coeval with the galaxy.}


\begin{table}
\caption{\label{color_poly} Integrated magnitude in the $g$ band and $g-r$ color derived in the polygons covering the substructures shown in Fig.~\ref{bmodel}.}
\begin{tabular}{lccc}
\hline\hline
region & structure&$m_g$ & $g-r$\\
         & & [mag]& [mag] \\
(1) & (2) & (3)& (4)\\
\hline
\hline
$I$ & L1& $12.328\pm 0.004$ & $0.634\pm0.006$\\
$II$ & L1& $13.53\pm 0.01$ & $0.63\pm0.01$\\
$III$ & L6& $14.48\pm 0.01$ & $0.58\pm0.02$\\
$IV$ & L7& $15.51\pm 0.02$ & $0.58\pm0.03$\\
$V$ & L2& $12.318\pm 0.008$ & $0.54\pm0.01$\\
$VI$ & L4& $13.346\pm 0.009$ & $0.65\pm0.01$\\
$VII$ & R1& $12.351\pm 0.003$ & $0.679\pm0.005$\\
$VIII$ & plume& $11.763\pm 0.003$ & $0.645\pm0.004$\\
$IX$ & &$11.096\pm 0.002$ & $0.672\pm0.004$\\
$X$ & centre& $11.906\pm 0.002$ & $0.860\pm0.003$\\
 & $L5$ & $12.88\pm 0.02$ & $0.65\pm0.03$\\
\hline
\end{tabular}
\tablecomments{{\it Col.1 -} and {\it Col.2 -} Number identifying the regions and features shown in Fig.~\ref{bmodel}. {\it Col.3 -} integrated magnitude in the $g$ band.  {\it Col.4 -} integrated colors. Values were corrected for the Galactic extinction by using the absorption coefficient $A_{g}=0.069$ and $A_{r}=0.048$ derived according to \citet{Schlegel98}.}
\end{table}


\section{Discussion: tracing the build-up history of Fornax~A}\label{discussion}

{   The deep and extended photometry of the VST mosaics allow us {\it i)} to trace with a great detail the morphology and structure of NGC~1316 out to  33~arcmin ($\sim 200$~kpc $\sim15R_e$) from the galaxy centre and down to $\mu_g \sim 31$~mag arcsec$^{-2}$  and $\mu_r \sim 29$~mag arcsec$^{-2}$; {\it ii)} to estimate the scales of the main components dominating the light distribution and {\it iii)} to give some constraints on the origin of the substructures found in each component by deriving the integrated colors.
The main results are listed below.

\begin{enumerate}

\item 
New faint loops are discovered on the N-NW side of the galaxy envelope (we named L6, L7, L8 and L9), in the range of surface brightness $29 \le \mu_r \le 30$~mag/arcsec$^{2}$ (see Fig.~\ref{zoomcenter} and Fig.~\ref{n1316mur}) and 
the giant ($\sim 0.44$~degree $\sim 160$~kpc) L5 loop ($\mu_g \sim 29 - 30$~mag arcsec$^{-2}$) in the SW region is confirmed. 
 The VST mosaic unveils the whole structure of L5: close to the galaxy it has a filament-like structure, while several bright blobs dominate the emission on the SE part of it. 
On the opposite side to L5, with respect to the galaxy centre, the new outer loop L9 on the NW is fainter and less extended than L5. It appears as a very curved arc with a radius of about 15~arcmin ($\sim 91$~kpc).
Inside the main body, the deep VST images reveal that  loops L1 and L4  are part of a unique and more extended structure,  they seem to be connected by a faint bridge (see Fig.~\ref{bmodel}). 

\item 
By fitting the azimuthally averaged surface brightness profile in the $r$ band, we identify two distinct components in NGC~1316.
Inside $R\le5.5$~arcmin ($\sim33$~kpc), i.e. for $R\leq2.5R_e$, the main body of the galaxy has a dominant $\sim r^{1/4}$ spheroidal component  (see Sec.~\ref{2Dmodel}). 
For $R\ge5.5$~arcmin, we map the outer envelope out to about 200~kpc from the galaxy centre. It has an  exponential decline (see Fig.~\ref{conf1399}) and a steep gradient towards bluer colors with respect to the central spheroid (see Fig.~\ref{colormap}). The presence of the two components is also consistent with the variations measured in the isophotes' ellipticity and P.A. profiles, in correspondence of the transition radius given above (see Fig.~\ref{epsPA}).  

\item 
 The loops L1, L2, L3, L4 and the new ones L6 and L7 are inside the stellar envelope. They are bluer than those in the central spheroid (see Fig.~\ref{zoomcenter}, Fig.~\ref{colormap} and Fig.~\ref{bmodel}). The bluest ones are L2 and L3, on the East side, with $g-r\sim0.54$~mag. The new loops L6 to L8, located on the N-NW side of the envelope, have comparable colors, $g-r=0.58$~mag, suggesting that they could be the brightest part of a single giant loop.
The giant SW loop L5, even if it is in the region of the envelope, has similar colors to those of the regions of the galaxy, which is $g-r\sim0.65$~mag (see Fig.~\ref{bmodel}). 

\end{enumerate}

The analysis of the integrated colors for the different structures in the central spheroid and outer envelope suggests that {\it i)} the giant loop L5 could be ``coeval'' with the central spheroid, since they have comparable colors, thus it can be considered as  a tidal tail of material expelled from the galaxy during a merging event in its formation history; {\it ii)} the  plume and  the extended tail detected SE to it (see  Fig.~\ref{colormap_zoom}), are the bluest features inside the central spheroid, 
therefore it is reasonable to assume that this structure was not formed together with the host galaxy, but it  could be the remnant of a smaller infalling galaxy; {\it iii)} the bluest loops detected in the galaxy envelope could be the sign of the minor accretion events that are contributing to the mass assembly of this component. 
Therefore, based on the above results, the main conclusion of this work  is that {\it for NGC~1316 we are still observing the relics (i.e. loop L5 and L9) of the first main process that formed the central spheroid and we are catching in act the second phase of the mass assembly \citep{Oser2010,Cooper2010}, where the accretion of smaller satellites is ongoing and it is building up the galaxy outskirts.}

In the following subsections we  discuss in detail the main formation processes that could have formed the central spheroid and the envelope by comparing the observed properties with the theoretical predictions. }

\subsection{  How did the central spheroid form?}

%

{  
To date, from the extensive literature available for NGC~1316 (see Sec.~\ref{n1316}), there are two processes proposed for the formation of this galaxy: the major merging of two disks galaxies and the minor merging of a giant early-type disk galaxy with a lower-mass late-type galaxy.

Numerical simulations of a major disk-disk merger show that during the final phase of such a merger, the violent relaxation transforms the stellar content of the two disks  into a spheroidal remnant having an $r^{1/4}$ density distribution \citep{Barnes1992,Hernquist1992,Mihos1996}. This would be consistent with the surface brightness distribution observed for NGC~1316 (see Sec.~\ref{fit}). Besides, the presence of shells or tidal tails also detected in this galaxy can be considered a "sign" of the past merging in the remnant object. 

Alternatively, the shells and ripples, as well as the nuclear dust, clearly detected in NGC~1316, could have formed throughout  the accretion of one, or more, smaller companion galaxies \citep[e.g.,][]{Quinn1984} by an early-type disk galaxy. According to the theoretical predictions, this formation process can work in forming tails with different length and surface brightness \citep{Balcells1997} as observed in NGC~1316, which are the SW loop L5 and the fainter and less extended loop L9 on the NW  (see Fig.~\ref{n1316mur}). If the accreted galaxy is massive enough, the giant loop L5 could be the returning material during the merging  \citep[e.g.,][]{Hernquist1992}. Taking into account that both loops are void of ionised gas \citep{Schweizer80,MF1998}, the brightest and reddest knots inside loop L5  (see Fig.~\ref{n1316mur} and Fig.~\ref{colormap}), are not star-forming regions, but more reasonably they could be remnant debris.

Based also on the main conclusion of this work, we suggest that the formation history of NGC~1316 is even more complex than those described above: after the main process that formed the central spheroid, many other accretion events are going on, which are building up the galaxy outskirt and are perturbing the inner structure. 
Therefore, in order to reproduce all the observed properties of NGC~1316 (i.e. morphology, gas content, kinematics) one needs a set of simulations of the interacting/merging galaxies, spanning a wide range of masses and morphological types, which include the secondary infall of small satellite galaxies. This is out of the scope of the present paper. 
However, as a starting point for a future work, we investigate how the main properties of the merger remnant formed by the two "simple" cases proposed above compare with the observations.
To this aim, we used  the {\it GalMer} simulations database \citep{Chilingarian2010}.
By looking at the remnants from all possible combinations of merging, with different mass ratio and morphological types, we extracted from the GalMer database the remnant at $\sim3$~Gyr, which is the upper limit age estimated for the last merger event \citep{Goudfrooij2001,Richtler2014},   that matches with  the observed properties in NGC~1316.
In particular, we check for the overall morphology, including the presence of shells/ripples/tails,  the surface brightness profile, the internal kinematics of stars and the amount of gas (see Sec.~\ref{n1316}). Results are discussed in the following section.}

\subsubsection{Tests on the formation scenarios for NGC~1316: disk-disk merging versus tidal accretion}

\begin{figure*}[t]
\centering
\includegraphics[scale=0.7]{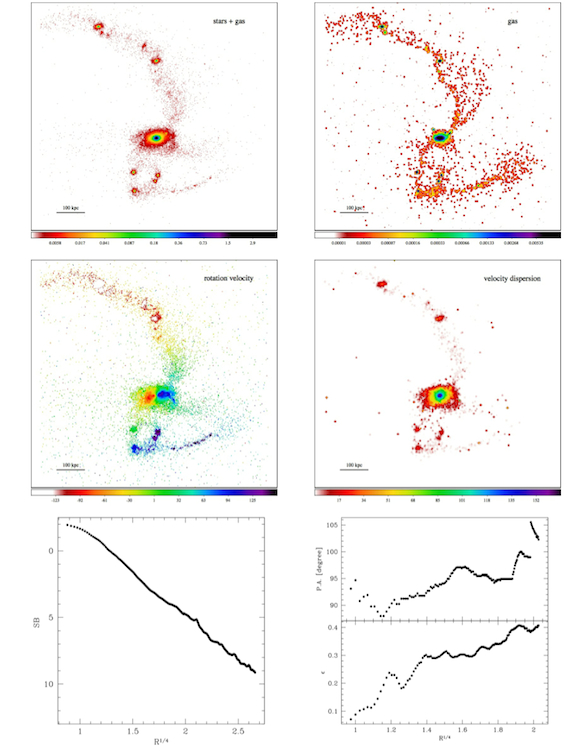}
\caption{\label{galmer1} The extracted snapshot at 3 Gyr from the GalMer database for a merging of two giant equal-mass disk galaxies. The total (stars + gas) surface density and the gas surface density are shown in the top left and top right panels, respectively. The stellar kinematics is shown in the middle panels (rotation velocity on the left and velocity dispersion on the right). The bottom panels show the azimuthally-averaged  surface brightness profile (left) and the P.A. and ellipticity profiles (right) of the fitted isophotes.}
\end{figure*}

\begin{figure*}[t]
\centering
\includegraphics[scale=0.7]{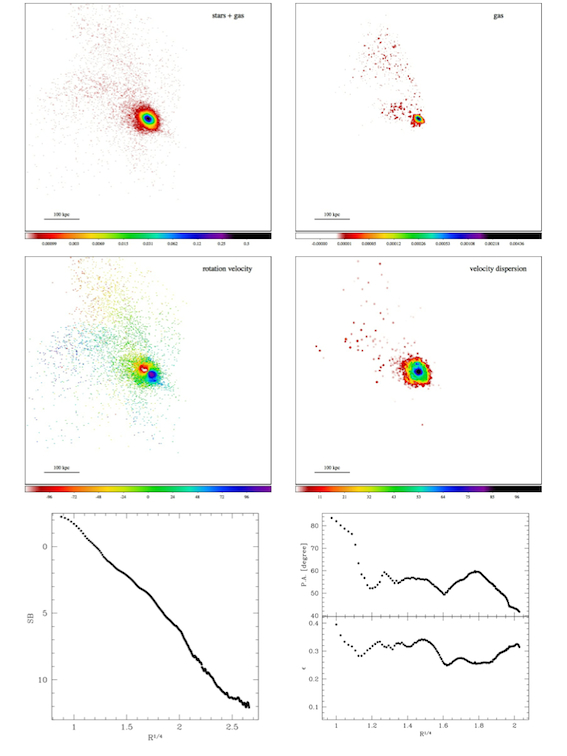}
\caption{\label{galmer2} The extracted snapshot at 3 Gyr from the GalMer database for a merging of two giant equal-mass disk galaxies. The total (stars + gas) surface density and the gas surface density are shown in the top left and top right panels, respectively. The stellar kinematics is shown in the middle panels (rotation velocity on the left and velocity dispersion on the right). The bottom panels show the azimuthally-averaged  surface brightness profile (left) and the P.A. and ellipticity profiles (right) of the fitted isophotes.}
\end{figure*}

 In Fig.~\ref{galmer1} and  Fig.~\ref{galmer2} we show the results from the GalMer database\footnote{From the GalMer database at http://galmer.obspm.fr/ the selected simulations have the following orbital parameters. Major merging: mass ratio=1:1; spin=prograde; inclination=45 deg; initial distance=100 kpc; pericentral distance = 8 kpc; motion energy = 2.5. The snapshot is extracted at the time of 3 Gyr. The angle of the viewport are $\phi=68$~deg and $\theta=-31$~deg. Minor merging: mass ratio=10:1; spin=prograde; inclination=33 deg; initial distance=100 kpc; pericentral distance = 8 kpc; motion energy = 2.5. The snapshot is extracted at the time of 2.5 Gyr. The angle of the viewport are $\phi=-96$~deg and $\theta=-51$~deg.} for a major merging of two disks and for a minor merging of a pre-existing giant S0 galaxy with a dwarf late-type disk, respectively. We show the density distribution of stars plus gas  (top-left panels), the distribution of gas (top-right panels) and the 2-dimensional map of the stellar  kinematics, i.e. rotation velocity and velocity dispersion (middle panels).
The above maps allow us to qualitatively describe the main expected structural properties of the remnants, as the presence of tidal features, the total mass and luminosity, surface brightness profiles and isophote shape. They are listed in the Tab.~\ref{galmer_simul}, which includes also those observed in NGC~1316. 
In the following we discuss how observations and models compare. 

\begin{table*}
\caption{\label{galmer_simul} Main parameters of the remnants derived by the GalMer simulation, compared with those observed in NGC~1316.}
\begin{tabular}{lccccccccc}
\hline
model & $M_{ratio}$ & morphology & gas/star & SB profile & $\epsilon$ & $V_{rot}^{max}$ & $\sigma_c$ & $\sigma$[$3R_e$] & $M_{tot}$\\
  & &  & & & & [km/s] &  [km/s] & [km/s]& $10^{11}$ M$_\odot$ \\
(1) & (2) & (3) & (4) & (5) &(6) &(7) & (8) & (9) & (10)\\
\hline\hline
 &&&&&&&&&\\
Sa+Sa & 1 & loops $R\sim190$kpc & $10^{-3} - 10^{-4}$ & $r^{1/4}$ & $0.1 - 0.4$ & $120-130$ & 180 & 91& 2\\
 &&&&&&&&&\\
S0+dSa & 10 & no loops & $10^{-3} - 10^{-4}$ & $r^{1/4}$ + exp & $\sim0.3$ & $120-130$ & 160 & 70 & 1.3\\
 &&&&&&&&&\\
\hline
 &&&&&&&&&\\
NGC 1316 & & loops $R\sim165$kpc & $10^{-4}$ & $r^{1/4}$ + exp & $\sim0.3$ & $150-170$ & 250 & 150 & $5.2-8.3$\\
\hline
\end{tabular}
\tablecomments{{\it Col.1 -}  GalMer simulations: {\it Sa+Sa} indicates the major merging of two equal-mass spiral galaxies; {\it S0+dSa} is the merging of a giant S0 galaxy with a dwarf late-type galaxy, which is 10 times less massive than the S0.  {\it Col.2 -}  Mass ratio of the merging galaxies. {\it Col.3 -} Features observed in the remnant like tails and loops. {\it Col.4 -} Mass ratio between the gas amount and mass of the stars. {\it Col.5 -} Shape of the azimuthally-averaged surface brightness profile (SB). {\it Col.6 -} Ellipticity of the isophotes. {\it Col.7 -} Maximum rotation velocity of the stars. {\it Col.8 -} and {\it Col.9 -} Central velocity dispersion of the stars and its estimate at $R_e$, respectively. {\it Col.10 -} Total stellar mass.
In the last line of the table are listed all the above quantities observed for NGC~1316. }
\end{table*}

{\it Morphology -} A spheroidal remnant galaxy with shells and extended tidal loop on the south, plus other faint tails in the outskirts (see left-top panel of Fig.~\ref{galmer1}), is obtained by an equal-mass merging of two giants spiral galaxies, with a total stellar mass of  $\sim10^{10}$~M$_{\odot}$ and a fraction of gas of about 0.1. 
In the adopted projection on the sky of the remnant, the south tidal loop extends out to about 190 kpc and has several bright knots. The NW tidal tail is less luminous than the southern one. For both features, from the density map shown in the top-left panels of Fig.~\ref{galmer1}, we derived the density ratio 
between the mass of the tails and the mass in the centre: it is  about $10^{-3}-10^{-5}$. In NGC~1316 the south tidal loop has a radius of about 160~kpc and it is characterised by several bright knots (see Fig.~\ref{n1316}), thus similar in shape and extension to that in the simulated merger remnant. Taking into account that the surface brightness of the south loop is in the range $29 \le \mu_r \le 30$~mag~arcsec$^{-2}$ (see Fig.~\ref{n1316mur}), 
this corresponds to $\sim10^{-5}$ fainter than the galaxy centre. Such a value is therefore comparable with that estimated in the merger remnant.

The minor merging of a pre-existing S0 galaxy with a dwarf late-type disk (mass ratio 10:1) also generates a spheroidal remnant galaxy, which has a very perturbed morphology by ripples and outer faint tails (see left-top panel of Fig.~\ref{galmer2}). Differently from the previous model, there are no prominent loops in the galaxy outskirts. 

{\it Gas content -} {  In the selected remnants, gas is concentrated at small radii, with some other fainter emissions associated to the outer tails (see the rigth-top panel of Fig.~\ref{galmer1}).} In both remnants, the ratio between the density of stars to that of the gas in the centre is about $10^{-3}-10^{-4}$ which is consistent with a ratio of about $10^{-4}$ observed in NGC~1316. Differently from what is observed in NGC~1316, gas is also detected in the outer tails formed in the disk-disk major merging. The absence of prominent star-forming tidal tails  and the lack of H{\small I} in them \citep{Horellou2001} suggest that NGC~1316 would be an evolved disk-disk merger remnant.

{\it Surface brightness and isophote shape -} By performing an ellipse fitting of the isophote in both remnants we found that, as observed in NGC~1316 (see Sec.~\ref{2Dmodel}), the azimuthally averaged surface brightness profile of the main body follows an $r^{1/4}$ behaviour, with several bumps at larger radii that are the sign of shells and tails (see the bottom panels of Fig.~\ref{galmer1} and Fig.~\ref{galmer2}). For the minor merger remnant, the surface brightness profile shows a change in the slope at larger radii, which reflects the existence of a second component. 
In both models the isophotes have an ellipticity (0.3-0.4) and twisting comparable to those found in NGC~1316 (see Fig.~\ref{epsPA}).
For the major merging remnant, the ellipticity increases with distance from the galaxy centre (from 0.1 to 0.4), while in the minor merging remnant it remains almost constant at about 0.3, more similar to what observed in NGC~1316.

{\it Kinematics -}  {  The stars in both remnants have a significant rotation with $V_{max} \sim 120 - 130$~km/s and quite large velocity dispersion in the centre ($\sigma_c \sim 160-180$~km/s, see Tab.~\ref{galmer_simul}).
The $V_{max}/\sigma_c$ is about 0.67 in the major merging remnant, which is more similar to the value of $\sim0.6$ observed in NGC~1316. The remnant by the other proposed scenario has a bit larger $V_{max}/\sigma_c$ ratio ($\sim 0.75$).
The main difference is in the velocity dispersion at larger radii, since for NGC~1316 it remains at higher levels ($\sim 150$~km/s at $3R_e$, see Tab.~\ref{galmer_simul}) than in the simulations. As we pointed out before,  in  NGC~1316 several secondary accretion events are contributing to the mass assembly, thus they could be responsible of the increasing velocity dispersion.}

{  In summary, the above analysis shows that the main difference between the two scenarios is the presence of the outer tails, observed in NGC~1316, which are expected for a major disk-disk merger but not so much for the unequal-mass merger. Therefore, taking also into account that the  $V_{max}/\sigma_c$ ratio is more similar to the value observed in NGC~1316 (see Tab.~\ref{galmer_simul}), the disk-disk major merging might be slightly favoured. In contrast to this, 
the extension, gas content and surface density of the outer tails do not fit with those observed in NGC~1316. In the model, the outer tails have comparable extensions and density (see Fig.~\ref{galmer1}). In NGC~1316 the NW loop L9 is less extended and fainter than the SW loop L5 (see Fig.~\ref{n1316mur}).} 
Also the gas content is different: in the tidal tails resulting from the disk-disk merging, the HI gas is detected along the whole extension of both tails, while the minor merging of an S0 with a dwarf late-type disk have a very small amount of neutral gas in the outer faint tails.
In NGC~1316, there is no detection of  neutral atomic gas in correspondence of the two giant outer loops L5 and L9, while emission is detected in four clumps in the galaxy outskirts {  (see Sec.~\ref{n1316}). 

As a conclusive remark, a more complex simulation is therefore needed to account for the structure and kinematics of NGC~1316. In particular, by studying  the faint galaxy outskirts, one key point is turned to be the structure and stellar population of the outer tails, which have different extension, are devoid of gas and star-forming regions and have comparable colors to the main body. These characteristics would exclude the major disk-disk merging as formation process for NGC~1316.
The unequal-mass merging of an early-type disk with a companion galaxy remains as probable scenario to account for the formation of NGC~1316. It also reconciles with results by dynamical models made for NGC~1316, which suggested that the dark matter halo  shows a central density comparable to those of massive elliptical galaxies \citep{McNeil2012,Richtler2014}, rather than of spiral galaxies. But it needs further investigations. One case to be verified is the interaction of a giant S0 galaxy with an intermediate late-type galaxy, to check if this process might give different results about the structure of the tidal tails. This option is not available in the GalMer database.}

\subsection{The building up of the stellar halo in NGC~1316}\label{halo}

In this section we attempt to trace the building up history of the stellar halo of NGC~1316 by comparing the observed properties with the theoretical predictions. 

{  Cosmological simulations of galactic stellar halo formation by the tidal disruption of accreted material are able to give a complete set of "observables", as structure, density profiles and metallicity, that can be compared with real data. }
%
 \citet{Cooper2010} made a detailed analysis of the properties for six simulated stellar haloes. They found that the assembly history  is made either by a gradual accretion of several progenitors with roughly equal mass  or by the accretion of one or two systems. 
The resulting morphology of the stellar halo is different in the two cases \citep[see Fig.6 in][]{Cooper2010}.  
The stellar haloes built up by the gradual accretion are the most extended, out to 70-100 kpc, and have several streams, shells  and other irregular structures, which are more evident at larger radii. On the contrary, the accretion of one or two massive satellites generates smaller stellar haloes with a strong central concentration. All features span a surface brightness range $24-35$~mag/arcsec$^2$ in the V-band.
The brightest and most coherent structures detectable in the stellar halo come from the most recent accretion events.
Also the shape of the density profiles changes  between the haloes formed by many progenitors and those formed by  few progenitors, being the former steeper and well fitted by a Sersic profile with $n\sim1$.
Haloes with several dominant progenitors show an almost flat metallicity profile, out to the outermost radii, while stronger metallicity variations are observed in case of a small number of accreted satellites.

The deep VST photometry allows us to map the faint regions of the stellar envelope of NGC~1316 out to 33~arcmin ($\sim200$~kpc). {  This component starts to dominate the light for $R\geq5.5$~arcmin ($\sim33$~kpc) and has a surface brightness in the range  $26-30$~mag/arcsec$^2$ (see Sec.~\ref{fit}). Therefore,} 
at these large distances and surface brightness levels we are able to make a direct comparison between the observed properties and those from theoretical predictions described above. 

The stellar envelope in NGC~1316 is characterised by several discrete and luminous substructures: the south extension of the loop L1, the whole bright part of the loops L2 and L3, and all the new fainter loops L6, L7 and L8 detected in the VST images (see Fig.~\ref{bmodel}, top panel). All of them have bluer $g-r$ colors than the substructures in the inner spheroid (see Fig.~\ref{bmodel}, bottom panel). At larger radii (for $R\ge20$~arcmin), the light in the stellar envelope appears  in a more diffuse form, which is probably mixed with the unbound component of the intracluster light. Probably, many other fainter loops are still present but their luminosities are below our detection limit.
Compared to the simulated haloes presented by \citet{Cooper2010}, the stellar halo in NGC1316 has very similar morphology, surface brightness levels, light distribution to those of  the Aq-C model (shown in Fig~6 of that paper), 
even if the simulated galaxy is about one order of magnitude less massive than NGC~1316. 
The Aq-C model is characterised by several loops in the range of V-band surface brightness $27-31$~mag/arcsec$^2$, which corresponds to an r-band magnitude of $26.6-30.6$~mag/arcsec$^2$, fully consistent with the range of surface brightnesses in the halo of NGC~1316 ($26-30$~mag/arcsec$^2$). 
The Aq-C halo, would result from the  gradual accretion of many progenitors and it is characterised by an exponential light profile, as also found in NGC~1316 (see Sec.~\ref{fit}).
One of the most luminous substructure in the stellar envelope of NGC~1316 includes the loops L2 and L3, which should be considered as the most recent accreted satellite. It has $g-r = 0.54\pm0.01$~mag (see Tab.~\ref{color_poly}) that is comparable with those estimated for the  galaxies inside the stellar envelope (see Fig.~\ref{mosaic}), which are $0.54 \leq g-r \leq 0.60$~mag (see Tab.~\ref{mag_galaxies}). Its total luminosity in the $g$ band, $L\ge 7 \times 10^9$~L$_{\odot}$, is comparable to those of the two brightest galaxies inside the envelope, NGC~1310 with $3 \times 10^9$~L$_{\odot}$ and NGC~1316C with $10 \times 10^9$~L$_{\odot}$  (see Fig.~\ref{mosaic} and Tab.~\ref{mag_galaxies}).
%
{  The $g-r$ colour gradient observed in the stellar envelope of NGC~1316 (see Fig.~\ref{colormap}) can also be an indication of a metallicity gradient, also consistent with the stellar halo formed by the accretion of small satellites.}

The fraction of the total stellar mass in surviving satellites in Aq-C model is $f_{sat}=0.28$, which contribute for about the 67\% to the total mass of the stellar halo  \citep[see Tab.2 in][]{Cooper2010}. In NGC~1316 we can estimate a lower limit for such a fraction, due to the detection limit of the VST photometry. {  This could be given by at least all loops clearly detected in the envelope (L1, L2, L3, L6 and L7)}. The total luminosity of these survived satellites remnants,  given by the sum of each single contribution (see Tab.~\ref{color_poly}), is $\sim 2  \times 10^9$~L$_{\odot}$ in the $g$ band. This is about 7\% of the total luminosity of NGC~1316  (see Sec..~\ref{ellipse}) and about 13\% of the total luminosity of the stellar envelope (see Tab.~\ref{param_fit}).

\subsection{NGC~1316 versus NGC~1399: two giants in comparison}

NGC~1316 is the giant spheroidal galaxy dominating the sub-group in the south-west of the Fornax cluster \citep{Drinkwater2001}, with an absolute magnitude in the $g$ band of $M_g=-23.16$~mag (see Sec.~\ref{ellipse}). {  It is  brighter than NGC~1399, the central cD galaxy of the Fornax cluster, which has   $M_g=-22.93$~mag \citep{Iodice2016}.}

The light profiles of the two giant galaxies have a very similar shape (see Fig.~\ref{conf1399}). Except for the difference in the luminosity, the inner parts for both galaxies are well fitted with a de Vaucouleurs law (see Sec.~\ref{fit}). The outer envelope is characterised by an exponential decline, with comparable level of central surface brightness and scale lengths.
In contrast to the very similar surface brightness profiles, the 2D morphologies of NGC~1316 and NGC~1399 are very different.
As illustrated in the present paper, as well as discussed in the previous works (see Sec.~\ref{n1316}), NGC~1316 is a dusty spheroid, showing evidence for past major and minor interactions, with a significant rotation of stars along its major axis \citep{Bedregal2006,Arnaboldi1998}. 
On the other hand, NGC~1399 appears as a "normal", slow rotator, elliptical galaxy, without dust and any sign of recent major gravitational interaction  \citep[see][and references therein]{Iodice2016}. 

The stellar envelope of NGC~1399 contributes more than the 60\% to the total light and it is in a very diffuse form. In NGC~1316 this component is a $\sim40\%$ fraction of the total light (see Sec.~\ref{fit}), where several discrete structures, remnants of the accreted progenitors,  are still visible at the same fainter level of surface brightnesses of the stellar envelope in NGC~1399. 
In both galaxies, the stellar envelope reaches the intracluster region, including some cluster galaxy members. Most of the galaxies inside the envelope of NGC~1316 are candidates dwarf galaxies, with $0.54 \leq g-r \leq 0.6$~mag (i.e.  $0.76 \leq V-I \leq 0.79$~mag)
plus two brighter barred spiral galaxies, NGC~1310 and NGC~1317 (see Fig.~\ref{mosaic} and Tab.~\ref{mag_galaxies}). 
On the contrary,  the stellar envelope of NGC~1399 hosts several bright early-type galaxies \citep{Iodice2016} and a vast populations of dwarfs, about 65 objects \citep{Mieske2007,Munoz2015}, with average colours $0.5\leq V-I \leq 1.2$~mag.

{  Close to the centre ($R\le0.1$~arcmin), due to the strong dust absorption NGC~1316 is slightly redder ($g-r \sim0.85$~mag, see Fig.~\ref{colormap}) than NGC~1399 \citep[$g-r \sim0.8$~mag][see]{Iodice2016}. The main body of both galaxies, where the light profiles are well fitted by a de Vaucouleurs' law (see Sec.~\ref{fit}), has similar $g-r$ color, $\sim 0.7-0.8$~mag. On the other hand, in NGC~1316 the outer envelope shows a gradient towards bluer colors, with $0.7 \leq g-r \leq 0.4$~mag (see Fig.~\ref{colormap}), while in NGC~1399 the stellar halo is  redder, with  $g-r \sim 0.8$~mag.}

In the framework of the mass assembly within galaxy clusters, the above analysis suggests that NGC~1316 and NGC~1399 trace a  different epoch of the formation history for a cD galaxy.
In NGC~1316 the central $r^{1/4}$ spheroid and the outer stellar envelope are still forming. 
NGC~1399 is in a more evolved phase than NGC~1316, but the similarity in the average light distribution and colors between the galaxies may suggest that the central spheroid in NGC~1399 formed by similar processes supposed for NGC~1316, but in an earlier epoch.
This is also a further indication that in the formation history of NGC~1316 the pre-merger galaxy could be an early-type object.  

{  
In the stellar envelope of NGC~1316 the  remnant features of the accreted progenitors are still present, thus the accretion is ongoing, but the light distribution has already settled with an exponential decline. In NGC~1399 there are no signs of recent interactions, except the very faint bridge of light toward NGC~1387 \citep{Iodice2016}, but the stellar envelope has a comparable exponential decline (same magnitudes and scale length) as observed in NGC~1316 (see Sec.\ref{fit}). Therefore, we could suppose a similar formation process for the stellar envelope in both galaxies, but at different epochs. At the present time, the stellar envelope in NGC~1399 is more virialized.}

\subsection{On a larger scale: the environment of Fornax~A versus the cluster core}

 The group of galaxies around Fornax~A, at about 1~Mpc to the South-West of the Fornax cluster, is bound to it and probably infalling toward the cluster centre \citep{Drinkwater2001}.  The Fornax core and the subgroup centred on Fornax~A may lie along a large filamentary structure of dark matter that is collapsing and flowing in toward a common centre. The uniform and large coverage of the Fornax cluster and its subgroup given by the FDS with VST will provide  an unprecedent opportunity to study the formation history on the cluster scale. 
Even if a detailed discussion of this subject is  out of the scope of this paper, it is worth commenting about how the environments around the two cDs galaxies (NGC~1399 and NGC~1316) compare, as a complementary discussion to that on the formation history for the two cDs presented in the previous section.

\begin{figure}[t]
\centering
\includegraphics[scale=0.45]{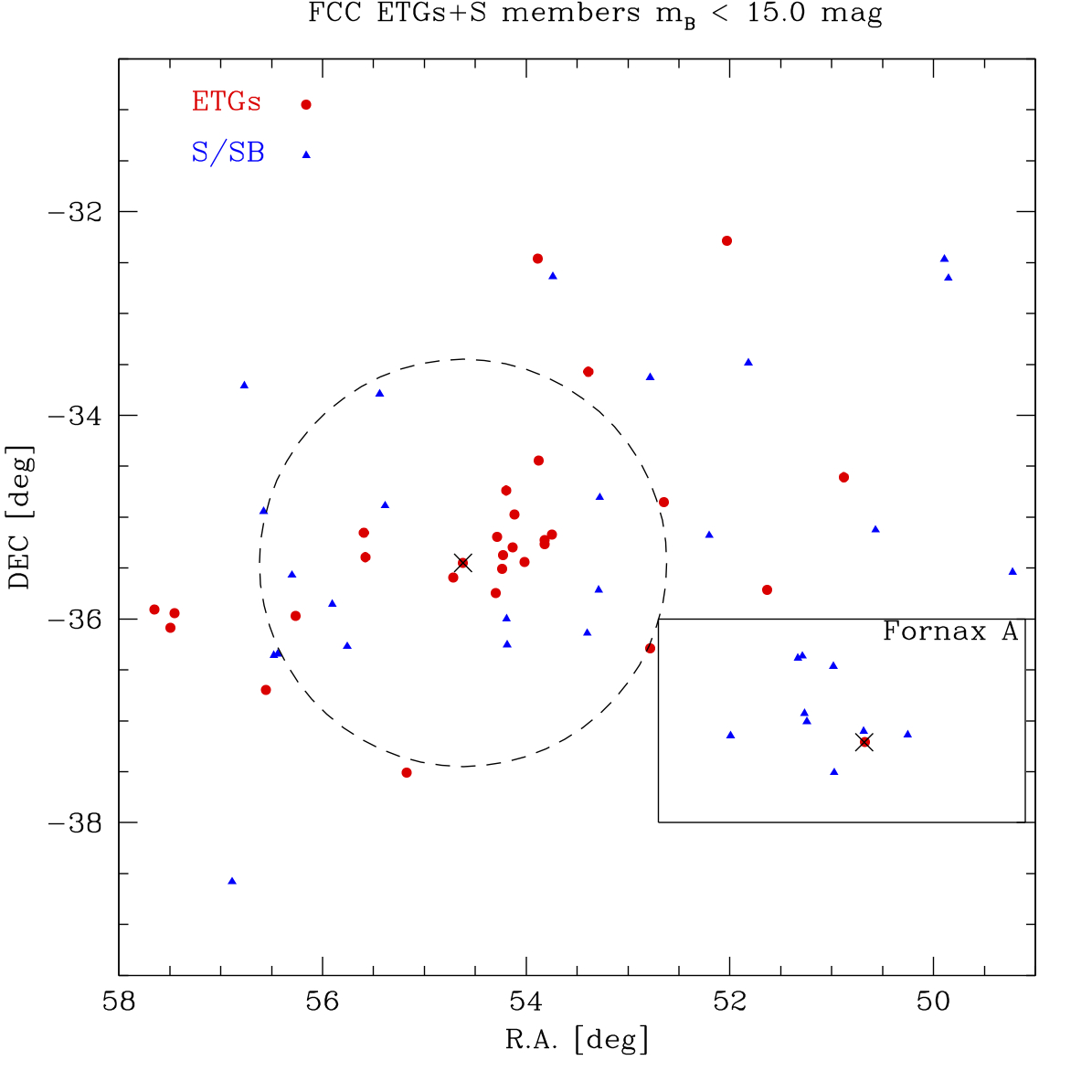}
\caption{\label{FCC} Distribution of early-type (Elliptical and S0s, red circles) and late-type galaxies (blue triangles) in the core of the Fornax cluster and in the sub-group Fornax~A. The two crosses indicate the location of NGC~1399 at the centre of the cluster and of NGC~1316 at the centre of the Fornax~A subgroup. The dashed circle corresponds to the virial radius $R_{vir}\sim0.7$~Mpc. The box delimits the region around Fornax~A studied in this work and presented in Fig.~\ref{mosaic}.}
\end{figure}

Fig.~\ref{FCC} shows all galaxies brighter than $m_B \le 15$~mag from the Fornax Cluster Catalogue by \citet{Ferguson1989}. Most of the galaxies inside the virial radius of the Fornax cluster are early-type galaxies, Ellipticals and S0s. The deep VST mosaic of the central 2 square degrees shows that they appear as "normal" spheroids, some of them are barred S0s, with no evident signs of interaction and/or accretion at the lowest magnitudes levels and star forming regions \citep{Iodice2016}. On the other hand,
in the 4 square degrees around NGC~1316 there are only late-types galaxies, barred spirals and interacting pairs, with luminous knots of active star forming regions (see Fig.~\ref{FCC} and Fig.~\ref{mosaic}). 
{  The differences in the environment around NGC~1316 and NGC~1399 are consistent with a different evolutionary epoch claimed above: the subgroup centred on NGC~1316 is in an earlier phase of mass assembly than the core of the cluster, which appears more evolved.   }

\section{Concluding summary}

As part of the Fornax Deep Survey with VST, we have obtained a new mosaic of the south-west group of the Fornax cluster, in the $g$ and $r$ bands, which covers an area of about  $3 \times 2$~square degrees around the central galaxy NGC~1316 (see Fig.~\ref{mosaic}). 
The deep photometry, the high spatial resolution of OmegaCam and the large covered area allow us, in one shot, to study the galaxy structure, to trace stellar halo formation and look at the galaxy environment. In particular, we are able to set the scales of the main components dominating the light distribution, which are the central spheroid and the stellar envelope,  and to study their properties.
We analysed with a great detail the numerous substructures that characterises both components, and the integrated colors derived for each of them turned out to be a new important constraints for the formation history of this fascinating galaxy.

NGC~1316 had a rich history of interaction events. There was a "major" gravitational interaction that formed the central spheroid: it could be a major merging of two equal-mass spiral galaxies or an unequal-mass merging of a pre-existing early-type disk with a late-type galaxy. We identified the giant SW loop L5 and the smaller and fainter new NW loop L9 as the relics of this process, being a tidal tail of  material expelled during the merging. This claim is based on the similarity between the integrated $g-r$ colors of these features with the average $g-r$ color of the central spheroid. We made a detailed comparison between the global properties observed for NGC~1316 (i.e. morphology, kinematics and gas content) with those from numerical simulations reproducing the above mergings.
{  The observed properties for the outer tails in NGC~1316 are not reproduced by the disk-disk major merging and the other proposed scenario (i.e. unequal-mass merging of a pre-existing early-type disk with a late-type galaxy) needs to be further investigated.} Moreover, we are catching on act the second phase of the mass assembly, where the accretion of smaller satellites is going on. They are building up the galaxy outskirt and are perturbing the inner structure. In particular, 
the faint envelope in NGC~1316 still hosts the remnants of the accreted satellite galaxies that are forming the stellar halo. Among them, the most luminous is on the NE of NGC~1316, identified as loops L2 and L3, which has luminosity and colors similar to those of the brightest galaxy members of this sub-group of the Fornax cluster, inside the envelope. By comparing with cosmological simulations of galactic stellar halo formation, we found that the structure, extensions, light distribution and mass fraction of the stellar envelope in NGC~1316 are consistent with the stellar haloes built up by the gradual accretion of small satellites.

By comparing NGC~1316 with the brightest galaxy in the core of the Fornax cluster, NGC~1399, we noticed that, even if the structure of the two giants is quite different, the azimuthally-averaged surface brightness profiles are surprisingly similar. For both galaxies, the central spheroid has  an $\sim r^{1/4}$ behaviour and the outer envelope is characterised by an exponential decline, with comparable level of central surface brightness and scale lengths. We claimed that, in the framework of the mass assembly within galaxy clusters, NGC~1316 and NGC~1399 trace a  different epoch of the formation history for a cD galaxy.

On the cluster scale, the  environment around the two giant galaxies is quite different. In the 4 square degrees around NGC~1316 there are only late-types galaxies, with evident signs of interaction and star formations regions, while most of the galaxies inside the core of the Fornax cluster are ETGs. {  Based on this, we suggest  a different evolutionary epoch for the two regions of the cluster: the subgroup centred on NGC~1316 is in an earlier phase of mass assembly than the core of the cluster, which appears more evolved.}


\acknowledgments

This work is based on visitor mode observations taken at the ESO La Silla Paranal Observatory within the VST GTO Program ID 096.B-0582(A). 
{  The authors wish to thank the anonymous referee for his/her comments and suggestions  that allowed us to greatly improve  the paper. }
 Authors acknowledges financial support from the INAF VST funds and  wish to thank ESO for the financial contribution given for the visitor mode runs at the ESO La Silla Paranal Observatory. Enrichetta Iodice wish to thank the ESO staff of the Paranal Observatory for the support during the observations at VST. E.I. is also very grateful to T. Puzia and V. Antonuccio-Delogu  for the discussions and suggestions on the present work. TR acknowledges support from  the BASAL Centro de Astrof\'{\i}sica y Tecnologias Afines (CATA) PFB-06/2007. NRN and EI received support within PRIN INAF 2014 "Fornax Cluster Imaging and Spectroscopic Deep Survey". F-B., G. vdV and R.~P. acknowledge support from grant AYA2016-77237-C3-1-P from the Spanish Ministry of Economy and Competitiveness (MINECO).

\bibliography{fornax}



\end{document}